\documentclass[aps,prl,amssymb,12pt,floatfix]{revtex4}
\usepackage{subfigure}
\usepackage{graphicx}
\usepackage{rotating}
\usepackage{amsmath,amsthm}

\begin{document}
\title{Measuring the energy  landscape roughness and the transition state location of biomolecules using 
single molecule mechanical unfolding experiments}
\author{Changbong Hyeon$^1$ \& D. Thirumalai$^{1,2}$}
\affiliation{$^1$Biophysics Program, Institute for Physical Science and 
Technology, University of Maryland, College Park, MD 20742 \\
$^2$Department of Chemistry and Biochemistry, University of Maryland, College 
Park, MD 20742\\
\[\]
}

\date{\today}
\baselineskip = 22pt
\begin{abstract}
Single molecule mechanical unfolding experiments are beginning to 
provide profiles of the complex energy landscape of
biomolecules.  In order to obtain reliable estimates of the energy landscape
characteristics it is necessary to combine the experimental measurements 
(the force extension curves, the mechanical unfolding trajectories, force or loading rate dependent unfolding rates)  with 
sound theoretical models and simulations.  Here, we show how by using
temperature as a variable in mechanical unfolding of biomolecules in laser optical tweezer or AFM experiments the 
roughness of the energy landscape can be measured without making any assumptions about the underlying reaction coordinate.  The efficacy of
the formalism is illustrated by reviewing experimental results that have directly measured roughness in a protein-protein complex.  
The roughness model can also be used to interpret experiments on forced-unfolding of proteins in which temperature is varied. Estimates of other  aspects of the energy landscape such as free energy barriers or the transition state (TS) locations could depend on the precise model used to analyze the experimental data. We illustrate the inherent difficulties in obtaining the transition state location from loading rate or force-dependent unfolding rates.  Because the transition state moves  as the force or the loading rate is varied it is in general difficult to invert the experimental data unless the curvature
at the top of the one dimensional free energy profile is large, i.e the 
barrier is sharp.  The independence of the TS location on force holds good 
only for brittle or hard biomolecules whereas the TS location changes considerably 
if the molecule is soft or plastic. We also comment on the usefulness of extension of the molecule as a surrogate reaction coordinate especially in 
the context of force-quench refolding of proteins and RNA.       
\end{abstract}
\maketitle

\section{Introduction}
Representation of the large conformational space of RNA and proteins in terms
of (low-dimensional) energy landscape has played an important role in visualizing their folding routes \cite{DillNSB97,OnuchicCOSB04,HyeonBC05}.  It is suspected that the energy landscape 
of many evolved proteins is relatively smooth so that they can be navigated very efficiently.  By smooth we mean that the gradient of the energy landscape $\Delta F(\chi,
 R_g)$ towards the native basin of attraction (NBA) is "large" enough that the biomolecule does not get kinetically trapped  in  competing basins of
 attraction for long times during the folding process. 
Here, $F(\chi, R_g)$ is expressed in terms two non-unique variables, namely, 
the radius of gyration $R_g$ and $\chi$, (or equivalently $Q$ the fraction of native contacts), 
an order parameter that measures how similar a given conformation is to the native state.  However, perfectly smooth with energy landscapes are difficult to realize because of energetic and topological 
frustration \cite{ThirumAccChemRes96,ClementiJMB00}. 
In proteins, the hydrophobic residues prefer to be sequestered in the interior 
while polar and charged residues are better accommodated on the surfaces 
where they can interact with water. 
Often these conflicting requirements cannot be simultaneously satisfied and hence proteins can be energetically ``frustrated''. 
It is clear from this description that only evolved or well designed sequences can minimize 
energetic frustration. Even if a particular foldable sequence minimizes energetic conflicts it is 
nearly impossible to eliminate topological frustration which arises due to 
chain connectivity \cite{GuoBP95,ThirumalaiRNA00}. 
If the packing of locally formed structures is in conflict with the global
 fold then the polypeptide or polynucleotide chain is topologically frustrated. 
Both sources of frustration, energetic and topological, render the energy landscape rugged 
on length scales that are larger than those in which secondary structures ($\approx (1-2)$ $nm$) form even if folding can be globally described using only two-states (folded and unfolded).

An immediate consequence of frustration is that the free energy, projected 
along  a one dimensional coordinate, is rough on certain length scale and may be globally smooth on larger scale. 
Let us assume that the characteristic roughness $\overline{\epsilon}$ has a Gaussian distribution. 
Following footnote 45 in \cite{ThirumalaiPRA89} the overall transit time from the unfolded 
basin may be written as 
\begin{equation}
\tau_{U\rightarrow F}\approx \tau(\beta)\int^{\infty}_{k_BT}d\overline{\epsilon}
e^{\beta\overline{\epsilon}}e^{-\overline{\epsilon}^2/2\epsilon^2}\approx
\tau(\beta)e^{\beta^2\epsilon^2/2}
\label{eqn:roughness}
\end{equation}
where $\epsilon$ is the average value of ruggedness. 
The second part of the equation holds good at low temperatures. 
The additional slowing down in the folding time $\tau_{U\rightarrow F}$ 
arising from the second term in Eq.\ref{eqn:roughness}, was derived in an elegant paper by 
Zwanzig \cite{ZwanzigPNAS88} and was also obtained in \cite{Bryngelson89JPC} by analyzing the dynamics of Derrida's random energy model \cite{Derrida81PRB}. 
If $\beta\epsilon$ is small then $\tau_{U\rightarrow F}\approx \tau_0e^{\beta\Delta F^{\ddagger}}$ where $\Delta F^{\ddagger}$ is the overall folding 
free energy barrier. 

If folding takes place in a rough energy landscape then the characteristic 
scale-dependent time scale may be estimated as 
\begin{equation}
\tau(l)\approx\left\{ \begin{array}{ll}
     \tau_{SS}\approx (10-100) ns& \mbox{$l\approx (1-2)$ $nm$}\\
     \tau_{U\rightarrow F}& \mbox{$l\approx L$}\end{array}\right.
\label{eqn:fluctuation}
\end{equation}
where $L$ is the effective contour length of the biomolecule. 
We have assumed that structures on $l\approx (1-2)$ $nm$ form in $\tau_{SS}\approx l^2/D$ where the diffusion constant is on the order of $(10^{-6}-10^{-7})$ $cm^2/sec$. 
The estimate of $\tau_{SS}$ is not inconsistent with the time needed to form 
$\alpha$-helices or $\beta$-hairpin especially given the crude physical picture.

With the possibility of manipulating biological molecules (Fig.\ref{RoughreviewFig1.fig}), one molecule at a time, using force it is becoming possible to probe the features of their energy landscape (such as roughness and the transition state location) that are not easily possible using conventional experiments. 
Such experiments, performed using Laser Optical Tweezers (LOTs) \cite{TinocoARBBS04,Ritort06JPHYS} or Atomic Force Microscopy (AFM) \cite{FernandezTIBS99}, have made it possible to mechanically unfold proteins \cite{MarquseeScience05,FernandezNature99,RiefPNAS04,GaubSCI97,BustamantePNAS00,Oberhauser98Nature,BustamanteSCI94}, RNA \cite{Bustamante01Science,Bustamante03Science,Woodside06PNAS,Block06Science,TinocoBJ06,Tinoco06PNAS,OnoaCOSB04}, and their complexes \cite{Moy94Science,Fritz98PNAS,EvansNature99,Schwesinger00PNAS,ZhuNature03,NevoNSB03}, or initiate refolding of proteins \cite{FernandezSCI04} and RNA \cite{TinocoBJ06,Tinoco06PNAS}. 
These remarkable experiments show how the initial conditions affect refolding and also enable us to examine the response of biological molecules over a range of forces and loading rates. In addition, fundamental aspects of statistical mechanics, including non-equilibrium work theorems \cite{JarzynskiPRL97,Crooks99PRE}, can be rigorously tested using the single molecule experiments \cite{Bustamante02Science,Trepagnier04PNAS}. Here, we are concerned with using the data and theoretical models to extract key characteristics of the energy landscape of biological systems.

The crude physical picture of folding in a rough energy landscape (Fig.\ref{landscape}) is not meaningful unless the ideas can be validated experimentally which requires direct measurement of the roughness energy scale $\epsilon$, barrier height, etc.
In conventional experiments, in which folding is triggered by temperature, 
it is difficult to measure $\tau_0$ and $\Delta F^{\ddagger}$ even when $\beta\epsilon\equiv 0$ \cite{GruebeleNature03}. 
We proposed, using theoretical methods, that $\beta\epsilon$ can be directly measured using forced-unfolding of biomolecules and biomolecular 
complexes. 
The Hyeon-Thirumalai (HT) theory \cite{HyeonPNAS03}, which was based on Zwanzig's treatment \cite{ZwanzigPNAS88}, showed that if unbinding or unfolding lifetime (or rates) are known as a function of the stretching force ($f$) and temperature ($T$) then $\epsilon$ can be inferred without explicit knowledge of $\tau_0$ or $\Delta F^{\ddagger}$. Recently, the loading-rate dependent unbinding times of a protein-protein complex using atomic force microscopy (AFM) at various temperatures have been used to obtain an estimate of $\epsilon$ \cite{ReichEMBOrep05}. Similarly, variations in the forced-unfolding rates as a function of temperature of \emph{Dictyostelium discoideum filamin} (ddFLN4) were used to
estimate $\epsilon$ \cite{RiefJMB05}. Although alternate interpretation of the data is proposed for temperature effect on ddFLN4 the variation in unbinding or unfolding rates of proteins as a function of $f$ and $T$ provides an opportunity to obtain quantitative estimates of the energy landscape characteristics.

Single molecule force spectroscopy can also be used to measure force-dependent unfolding rates from which the location of the transition state (TS) in terms of the spatial extension ($R$) can be computed.  
This procedure is a not straightforward because, as shown in a number of studies \cite{HyeonBJ06,HyeonPNAS05,Lacks05BJ,West06BJ,AjdariBJ04}, the location of the transition state changes as $f$ changes unless the curvature of the free energy profile at the TS location is large, i.e, the barrier is sharp. 
The extent to which the TS changes depends on the load.  
We propose distinct scenarios for variation of the TS location as the external conditions change. 
By carefully considering the variations of force distributions it is possible 
to obtain reliable estimates of the TS location \cite{RiefJMB05}. 
Here, we review recent developments in single molecule force spectroscopy that have attempted to obtain the energy landscape characteristics of biological molecules \cite{BellSCI78,Evans97BJ,HyeonPNAS03,Dudko03PNAS,HummerBJ03,Barsegov05PRL,Barsegov06BJ,AjdariBJ04}. Using theoretical models we also  point out some of the ambiguities in interpreting the experimental data from dynamic force spectroscopy.

\section{Theoretical background}
Single molecule mechanical unfolding experiments differ from conventional 
unfolding experiments in which unfolding (or folding) is triggered by varying temperature or concentration of denaturants or ions. 
In single molecule experiments folding or unfolding can be initiated by precisely manipulating the initial conditions. 
In both forced-unfolding and force-quench refolding the initial conformation, characterized by the extension of the biomolecules, are precisely known. On the other hand, the nature of the unfolded states, from which refolding is initiated, is hard to describe in ensemble experiments.  
For RNA and proteins, whose energy landscape is complex \cite{TreiberCOSB01,HyeonBC05}, details of the folding pathways can be directly monitored by probing the time dependent changes in the end-to-end distance $R(t)$ of individual molecules.  
Analysis of such mechanical folding and unfolding trajectories allows to explore regions of the energy landscape that are difficult to probe using ensemble experiments.

Force experiments can measure the extension of the molecule as a function of time.  
There are three modes in which the stretching experiments are performed. 
Most of the initial experiments were performed by unfolding biomolecules (especially proteins) by pulling on one of the molecule at constant velocity while keeping the other end fixed \cite{Bustamante01Science,Bustamante03Science,FernandezNature99,Reif,Fisher00NSB}. 
More recently, it has become possible to apply constant force on the molecule of interest using feed-back 
mechanism \cite{Visscher99Science,Schlierf04PNAS,FernandezSCI04,TinocoBJ06,Fernandez06NaturePhysics}. 
In addition, force-quench experiments have been reported in which the forces are decreased or increased linearly \cite{FernandezSCI04,TinocoBJ06,Tinoco06PNAS}. 
It is hoped that a combination of such experiments can provide a detailed picture of the complex energy landscape of proteins and RNA.  
In all the modes, the  variable conjugate to $f$ is the natural coordinate that describes 
the progress of the reaction of interest (folding, unbinding or catalysis).
If there is a energy barrier confining the molecular motion to a local minimum, 
whose height is greater than $k_BT$, then a sudden increase (decrease) of extension (force) signifies the transition of the molecule over the barrier. 
A cusp  in the force-extension curve (FEC) is the signature of such a transition.   Surprisingly, for proteins and RNA it has been found that the
 FECs can be quantitatively fit using the semi-flexible or worm-like chain model \cite{MarkoMacro96,Bustamante97Science,Bustamante01Science,Bustamante03Science,FernandezNature99,Fisher00NSB}. 
From such fits, the global polymeric properties of the biomolecule, 
such as the contour length and the persistence length can be extracted \cite{MarkoMacro96,BustamanteSCI94}.

Single molecule experiments provide distributions of  the unfolding times (or unfolding force) by varying external conditions.  The objective is to construct the underlying energy landscape from such measurements and from mechanical folding or unfolding trajectories.  However, it is difficult to construct from FEC or mechanical folding trajectories that report only changes at two points all the features of the energy landscape of biomolecules.  For example, although signature of roughness in the energy landscape may be reflected as fluctuations in the dynamical trajectory  it is difficult to estimate its value unless multiple pulling experiments are performed.  We had proposed that the power of single molecules can be more fully realized if temperature ($T$) is also used as an additional variable \cite{Klimov01JPCB,Klimov00PNAS,HyeonPNAS03,HyeonPNAS05}. By using $T$ and $f$ it is possible to obtain the phase diagram as a function of $T$ and $f$  that can be used to probe the nature of collapsed molten globules which are invariably populated but are hard to detect in conventional experiments.  We also showed theoretically that the roughness energy scale ($\epsilon$) can be measured \cite{HyeonPNAS03} if both $f$ and temperature ($T$) are varied in single molecule  experiments.   The effect of $\epsilon$ manifests itself in the  $1/T^2$ dependence of the rates of force-induced unbinding or unfolding kinetics.  In the following subsections we first review the theoretical framework to describe the force-induced unfolding kinetics and show how the 
$1/T^2$ dependence emerges when the roughness is treated as a perturbation in the underlying energy profile.

{\it Bell model: }
Historically,  a phenomenological  description of the forced-unbinding of adhesive contacts was made by Bell \cite{BellSCI78} long before single molecule experiments were performed. In the context of ligand unbinding from binding pocket, 
Bell \cite{BellSCI78} conjectured that the kinetics of bond rupture can  be described using a  modified Eyring rate theory \cite{EyringJCP35}, 
\begin{equation}
k=\kappa\frac{k_BT}{h}e^{-(E^{\ddagger}-\gamma f)/k_BT}
\label{eqn:Erying}
\end{equation} 
where $k_B$ is the Boltzmann constant, $T$ is the temperature, 
$h$ is the Planck constant, and $\kappa$ is the transmission coefficient.  
In the Bell description the activation barrier $E^{\ddagger}$ is reduced by a factor $\gamma\times f$ when the bond or the biomolecular complex is subject to  
external force $f$.  The parameter  $\gamma$ is  a characteristic length of the system 
under load and specifies the distance at which the molecule unfolds or the ligand unbinds. 
The prefactor $\frac{k_BT}{h}$ is the vibrational frequency of a
single bond. The Bell model shows that the unbinding rates increase when tension is applied to the molecule.
Although Bell's key conjecture, i.e., the reduction of activation barrier due to external force, is physically justified, 
the assumption that $\gamma$ does not depend on the load is in general not valid. In addition, 
because of the multidimensional nature of the energy landscape of biomolecules, 
there are multiple unfolding pathways which require modification of the Bell description of forced-unbinding.  It is an oversimplification to restrict the molecular response to the force merely to a  reduction in the free energy barrier.
Nevertheless, in the experimentally accessible range of loads the Bell model in conjunction
with Kramers' theory of escape from a potential well have been remarkably successful in fitting much of the data on forced-unfolding of biological molecules. 

{\it Mean first passage times:}
In order to go beyond the popular Bell model many attempts have been made to describe unbinding process as an escape from a free energy surface in the presence of force \cite{Evans97BJ,HummerBJ03,HyeonPNAS03,BarsegovPNAS05,Dudko06PRL}.  
This is traditionally achieved by a formal procedure that adapts the  Liouville equation that describes the time evolution of the  
probability density representing the molecular configuration on the phase space.

For the problem at hand, one can project the entire dynamics onto a single reaction coordinate 
provided the relaxation times of other degrees of freedom are shorter than the time scale associated with the presumed order parameter of interest \cite{Zwanzig60JCP,ZwanzigBook}.  In applications to force-spectroscopy, we assume that the the variable conjugate to $f$ is a reasonable approximation to the reaction coordinate. The probability density of the molecular configuration, $\rho(x,t|x_0)$, whose 
configuration is represented by order parameter $x$ at time $t$,  obeys the Fokker Planck equation.

\begin{equation}
\frac{\partial\rho(x,t|x_0)}{\partial t}=\mathcal{L}_{FP}(x)\rho(x,t|x_0)=\frac{\partial}{\partial x}D(x)\left(\frac{\partial}{\partial x}+\frac{1}{k_BT}\frac{dF(x)}{dx}\right)\rho(x,t|x_0), 
\label{eqn:FP}
\end{equation}
where $D(x)$ is the position-dependent diffusion coefficient, 
and  $F(x)$ is an effective one-dimensional free energy, $x_0$ is the position at time $t=0$. 
If the initial distribution is given by $\rho(x_0,t=0|x_0)=\delta(x-x_0)$ 
the formal solution of the above equation reads $\rho(x,t|x_0)=e^{t\mathcal{L}_{FP}}\delta(x-x_0)$. 
If we use absorbing boundary condition at a suitably defined location, the 
probability that the molecule remains bound (survival probability)  at time $t$ is
\begin{equation}
S(x_0,t)=\int dx\rho(x,t|x_0)=\int dxe^{t\mathcal{L}_{FP}}\delta(x-x_0). 
\label{eq:S}
\end{equation}
In terms of the the first passage time distribution, $p_{FP}(x_0,t)(=-dS(x_0,t)/dt)$, the
mean first passage time can be computed using,
\begin{eqnarray}
\tau(x_0)&=&\int_0^{\infty}dt \left[t p_{FP}(x_0,t)\right]\nonumber\\
&=&-\int^{\infty}_0dt\left[t\frac{dS(x_0,t)}{dt}\right]\nonumber\\
&=&\int^{\infty}_0dt\int dxe^{t\mathcal{L}_{FP}(x)}\delta(x-x_0)\nonumber\\
&=&\int^{\infty}_0dt\int dx\delta(x-x_0)e^{t\mathcal{L}^{\dagger}_{FP}(x)
}
\label{eq:detail}
\end{eqnarray}
where  $\mathcal{L}^{\dagger}_{FP}$ is the adjoint operator. 
In obtaining the above equation we used $S(x_0,t=\infty)=0$ and integrated by parts in going from the second to third third line.
By operating on both sides of Eq.\ref{eq:detail} with $\mathcal{L}_{FP}^{\dagger}(x_0)$ and exchanging the variable $x$ with $x_0$ we obtain
\begin{equation}
\mathcal{L}^{\dagger}_{FP}(x)\tau(x)=e^{F(x)/k_BT}\frac{\partial}{\partial x}D(x)e^{-F(x)/k_BT}\frac{\partial}{\partial x}\tau(x)=-1. 
\end{equation}
The rate process with reflecting boundary $\partial_x\tau(a)=0$ and absorbing boundary condition $\tau(b)=0$ in the interval $a\leq x\leq b$, leads to 
the expression of mean first passage time, 
\begin{equation}
\tau(x)=\int^b_x dye^{F(y)/k_BT}\frac{1}{D(y)}\int^y_a dz e^{-F(z)/k_BT}.
\label{eqn:mfpt}
\end{equation}

{\it Diffusion in a rough potential:}
In the above analysis the one-dimensional free energy profile $F(x)$  that 
approximately describes the unfolding or unbinding event is arbitrary.  In 
order to explicitly examine the role of the energy landscape ruggedness we 
follow Zwanzig and  decompose $F(x)$ into $F(x)=F_0(x)+F_1(x)$ \cite{ZwanzigPNAS88}. 
where $F_0(x)$ is a smooth potential that determines  the global shape of the energy landscape, and $F_1(x)$ is the periodic ruggedness that superimposes $F_1(x)$. 
By taking the spatial average over $F_1(x)$  using 
$\langle e^{\pm\beta F_1(x)}\rangle_l=\frac{1}{l}\int^l_0dx e^{\pm\beta F_1(x)}$, where $l$ is the ruggedness length scale, the 
 associated mean first passage time can be written in terms of the effective diffusion constant $D^*(x)$ as, 
\begin{eqnarray}
D^*(x)&=&\frac{D(x)}{\langle e^{\beta F_1(x)}\rangle_l\langle e^{-\beta F_1(x)}\rangle_l}, \nonumber\\
\tau(x)&\approx&\int^b_x dye^{F_0(y)/k_BT}\frac{1}{D^*(y)}\int^y_a dz e^{-F_0(z)/k_BT}.
\label{eq:MFPT}
\end{eqnarray}
 An inversion of roughness barrier, i.e., $F_1\leftrightarrow -F_1$ does not 
alter $D^*(x)$. In the presence of roughness $D^*(x)\leq D(x)$.
Depending on the distribution of roughness barrier, $D^*(x)$ can take various forms:

\begin{enumerate}
\item
 For $F_1(x)=\epsilon x/b$ $(0\leq x\leq b)$ and $F_1(x)=\epsilon(a-x)/(a-b)$ $(b\leq x\leq a)$ with $F_1(x)=F_1(x+a)$, $\langle e^{\beta F_1(x)}\rangle\langle e^{-\beta F_1(x)}\rangle=\left[\frac{\sinh{\beta\epsilon/2}}{\beta\epsilon/2}\right]^2$ \cite{Festa78PhysicaA}. 
\item
For $F_1(x)=\epsilon\cos{qx}$, $\langle e^{\beta F_1(x)}\rangle\langle e^{-\beta F_1(x)}\rangle=[I_0(\beta\epsilon)]^2$ 
\cite{ZwanzigPNAS88}. 
\item
If $P(F_1)$ is Gaussian with variance $\langle F_1^2\rangle=\epsilon^2$,
$\langle e^{\beta F_1(x)}\rangle\langle e^{-\beta F_1(x)}\rangle=e^{\beta^2\epsilon^2}$ \cite{ZwanzigPNAS88}.
\item
If $P(F_1)=P(-F_1)$, then $\langle e^{\pm\beta F_1(x)}\rangle_l=\int dF_1P(F_1)\left[1+\frac{1}{2!}\beta^2\epsilon^2+\frac{1}{4!}\beta^4\epsilon^4+\cdots\right]$. 
\end{enumerate}
\noindent In all cases when $\beta\epsilon$ is small the effective diffusion coefficient can be approximated as $D_0^*\approx D\exp{\left(-\beta^2\epsilon^2\right)}$ where $D_0$ is the bare diffusion constant. If $P(F_1)$ is a Gaussian then this expression is exact.

From the recent experimental analysis on mechanical unfolding kinetics of the 
multi-ubiquitin construct, Fernandez and coworkers \cite{Fernandez06NaturePhysics} obtained a
power-law distribution of unfolding times i.e.,  $P(\tau)\propto \tau^{-(1+a)}$, and 
showed that the distribution of energy barrier heights should be $P(F_1)\sim\exp(-|F_1|/\overline{F_1})$ where $\overline{F_1}=k_BT/a$ \cite{Fernandez06NaturePhysics}. 
This case belongs to class 4, thus one obtains $e^{-\epsilon^2/k_B^2T^2}$ behavior associated with the effective diffusion coefficient.  These examples show that the coefficient associated with $1/T^2$ behavior is due to the energy landscape roughness provided the extension is a good reaction coordinate.
The dependence of $e^{-\epsilon^2/k_B^2T^2}$ in $D^*$ suggests that the diffusion in rough potential can be substantially slowed even when the scale of roughness is not too large.

{\it Barrier crossing dynamics in a tilted potential:} In writing the one-dimensional Fokker-Planck equation (Eq.\ref{eqn:FP}) we 
assumed that the order parameter $x$ is a slowly changing variable. 
This assumption is valid if the molecular extension, in the presence of $f$, describes accurately the conformational changes  in the biomolecule.

Following the Bell's conjecture we can replace $F(x)$ by $F(x)-f\cdot x$ in which $f$ "tilts" the free energy surface.  Thus,  in the presence of mechanical force Eq.\ref{eq:MFPT} becomes 
\begin{equation}
k^{-1}(f)=\tau(f;x)\approx\int^b_x dye^{(F_0(y)-fy)/k_BT}\frac{1}{D^*(y)}\int^y_a dz e^{-(F_0(z)-fz)/k_BT}. 
\label{eqn:MFPT_force}
\end{equation}
As long as the energy barrier is large enough (see Fig.\ref{landscape})
Eq.\ref{eqn:MFPT_force} can be further simplified using the saddle point approximation. 
The Taylor expansions of the free energy potential $F_0(x)-fx$ at the barrier top and the minimum result in the Kramers' equation \cite{KramersPhysica40,Hanggi90RMP}, 
\begin{eqnarray}
k^{-1}(f)=\tau(f)&\approx&\frac{2\pi k_BT}{D^*m\omega_b(f)\omega_{ts}(f)}e^{\beta(\Delta F_0^{\ddagger}(f)-f\Delta x(f))}\nonumber\\
&=&\left(\frac{2\pi\zeta}{\omega_b(f)\omega_{ts}(f)}\langle e^{\beta F_1(x)}\rangle_l\langle e^{-\beta F_1(x)}\rangle_l\right)e^{\beta(\Delta F_0^{\ddagger}(f)-f\Delta x(f))}
\label{eqn:Kramers}
\end{eqnarray}
where $\omega_b$ and  $\omega_{ts}$ are the curvatures of the potential, $|\partial^2_xF_0(x)|$, at $x=x_b$ and $x_{ts}$, respectively, the free energy barrier
$\Delta F_0^{\ddagger}=F_0(x_{ts})-F_0(x_b)$, $m$ is the effective mass of the biomolecule,
$\zeta$ is the friction coefficient, and $\Delta x\equiv x_{ts}-x_b$.

In the presence of $f$, the positions of transition state $x_{ts}$ and bound state $x_b$ change because unbinding
kinetics should be  determined using $F_0(x)-fx$ and not $F_0(x)$ alone. Because 
$x_{ts}$ and $x_b$ satisfy the \emph{force dependent condition} $F_0'(x)-f=0$, 
it follows that all the parameters, $\Delta x(f)$, $\omega_{ts}(f)$, and $\omega_b(f)$,  are intrinsically $f$-dependent. 
Depending on the shape of free energy potential $F_0(x)$, the degree of 
force-dependence of $\Delta x$, $\omega_{ts}$, and $\omega_b$ can vary greatly. 
Previous theoretical studies \cite{HummerBJ03,Dudko03PNAS,HyeonBJ06,RitortPRL06} have examined some of the consequences of the moving transition state.  
In addition, simulational studies \cite{HyeonPNAS05,Lacks05BJ,HyeonBJ06} in which  the free energy profiles  were explicitly computed from thermodynamic considerations alone clearly showed the 
change of $\Delta x$ when $f$ is varied.  These authors also provided a structural basis for transition state movements in the case of unbinding of simple RNA hairpins. 
The nontrivial coupling of force and free energy profile makes it difficult to 
unambiguously extract free energy profiles from experimental data. 
In order to circumvent some of the problems Schlierf and Rief have used Eq.\ref{eqn:MFPT_force}  to analyze the load-dependent experimental data on unfolding of ddFLN4 and extracted an effective
one dimensional free energy surface $F(x)$ without making additional assumptions.  The results showed that the effective free energy profile is highly anharmonic near the transition state region \cite{Schlierf06BJ}.

{\it Forced-unfolding dynamics at constant loading rate: } Many single molecule experiments are conducted by ramping force over time \cite{Bustamante02Science,Bustamante03Science,FernandezNature99,FernandezTIBS99}. In this mode the load on the molecule or the complex increases with time. When the force increases beyond a threshold value, unbinding or bond-rupture occurs. Because of thermal fluctuations the unbinding events are stochastic and as a consequence one has to contend with the distribution of unbinding forces.  The time-dependent nature of the force makes the barrier crossing rate also dependent on $t$. 
For a single barrier crossing event with a time-dependent rate $k(t)$, 
the probability of the barrier crossing event being observed at time $t$ is 
$P(t)=k(t)S(t)$ where the survival probability, that the molecule remains 
folded, is given as $S(t)=\exp{\left(-\int ^t_0d\tau k(\tau)\right)}$.

When the molecule or complex is pulled at a  constant loading rate ($r_f$) 
the distribution ($P(f)$) of unfolding forces is asymmetric.  The most probable $r_f$-dependent unfolding force ($f^*$) is often used to determine  the TS location of the underlying energy landscape with the tacit assumption that the TS is stationary. When $r_f=df/dt$ is constant, the probability of observing an  unfolding event at force $f$ is written as, 
\begin{equation}
P(f)=\frac{1}{r_f}k(f)\exp{\left[-\int^f_0df'\frac{1}{r_f}k(f')\right]}. 
\label{eqn:force_distribution}
\end{equation} 
The most probable unfolding force is obtained from
$dP(f)/df|_{f=f^*}=0$, which leads to
\begin{eqnarray} 
f^*&=&\frac{k_BT}{\Delta x(f^*)}\lbrace\log{\left(\frac{r_f\Delta x(f^*)}{\nu_D(f^*)e^{-\beta\Delta F^{\ddagger}_0(f^*)}k_BT}\right)}\nonumber\\ 
&+&\log{\left(1+f^*\frac{\Delta x'(f^*)}{\Delta x(f^*)}-\frac{\left(\Delta F^{\ddagger}_0\right)'(f^*)}{\Delta x(f^*)}
+\frac{\nu'_D(f^*)}{\nu_D(f^*)}\frac{k_BT}{\Delta x(f^*)}\right)}\nonumber\\ 
&+&\log{\langle e^{\beta F_1}\rangle_l\langle e^{-\beta F_1}\rangle_l}\rbrace,
\label{eqn:most_force}
\end{eqnarray}
where $\Delta F^{\ddagger}_0\equiv F_0(x_{ts}(f))-F_0(x_0(f))$, $\prime$ 
denotes differentiation with respect to the argument, $\Delta x(f)\equiv x_{ts}(f)-x_0(f)$ is the distance between the transition state and the native state, and 
$\nu_D(f)\equiv \omega_o(f)\omega_{ts}(f)/2\pi\gamma$. 
Note that $\Delta F^{\ddagger}_0$, $\nu_D$, and $\Delta x$ depend on the value of $f$ \cite{HyeonPNAS03,Lacks05BJ,HyeonPNAS05}.  
Because $f^*$ changes with $r_f$, $\Delta x$ obtained from the data analysis should correspond to a value at a certain $f^*$, not a value that is extrapolated to $f^*=0$.  Indeed, the pronounced curvature in the plot of $f^*$ as a function of $\log{r_f}$  makes it difficult to obtain the characteristics of the underlying energy landscape using data from dynamic force-spectroscopy without a reliable theory or a model.
If $\Delta F^{\ddagger}_0$, $\nu_D$, and $\Delta x$ are relatively insensitive to variations in  force, the second term on the right-hand side of Eq.\ref{eqn:most_force} would vanish, leading to $f^*\propto (k_BT/\Delta x)\log{r_f}$ \cite{Evans97BJ}. 
If the loading rate, however, spans a wide range so that the force-dependence of $\Delta F^{\ddagger}_0$, $\nu_D$, and $\Delta x$ are manifested, then the resulting $f^*$ can substantially deviate from the linear dependence to the $\log{r_f}$.  Indeed, it has been shown that for a molecule or a complex known to have a single free energy barrier, the average rupture force $\overline{f}\approx (\log {r_f})^{\nu}$ where the effective exponent $\nu \le 1$.  The precise value of $\nu$ depends on the nature of the underlying potential and is best treated as an adjustable parameter \cite{Dudko06PRL}.

Provided $\Delta F^{\ddagger}_0$, $\nu_D$, and $\Delta x$ are assumed constant the applicability of Kramers' equation can be checked.   In principle, the use of Kramers' equation is justified if  the effective energy barrier is at least greater than the thermal energy $k_BT$ ($\Delta^{\ddagger}=\Delta F^{\ddagger}_0-f^*\Delta x>k_BT$) at the most probable unfolding force $f^*$. Substitution of $f^*$ from Eq.\ref{eqn:most_force}  leads to 
\begin{equation}
\Delta^{\ddagger}=-k_BT\log{\frac{r_f\Delta x}{\nu_Dk_BT}}. 
\end{equation}
The condition that  $\Delta^{\ddagger}>k_BT$ is satisfied  as long as $r_f<\nu_D\frac{k_BT}{e\Delta x}=r_f^c$. 
For the set of parameters, $k_BT=4.14$ $pN\cdot nm$, $\Delta x\sim 1$ $nm$, and $\nu_D\sim 10^6$ $s^{-1}$, the critical loading rate $r_f^c\sim 10^6$ $pN/s$. 
The typical loading rate used in force experiments is several orders of magnitude smaller 
than this value. Therefore, it is legitimate to interpret the force experiments using the formalism based on the Kramers' equation. 
If $r_f>r_f^c$, as is typically the case in  steered molecular dynamics simulations \cite{SchultenBJ98}, the forced-unfolding process can no longer be considered a thermally activated barrier crossing process.  At such high loading rates average (not the same as $f^*$) rupture force ($\overline{f}$) grows (non-logarithmically) as $v^{1/2}$ where  $v$ is the pulling speed \cite{HummerBJ03}.

\section{Measurement of energy landscape roughness}
In the presence of roughness we expect that the unfolding kinetics deviates substantially from an Arrhenius behavior.
By either assuming a Gaussian distribution of the roughness contribution $F_1$  ($P(F_1)\propto e^{-F_1^2/2\epsilon^2}$) or simply assuming $\beta F_1\ll 1$ and $\langle F_1\rangle=0$, $\langle F_1^2\rangle=\epsilon^2$, one can further simplify 
Eq.\ref{eqn:Kramers} to 
\begin{equation}
\log{k(f)/k_0}=-(\Delta F_0^{\ddagger}-f\cdot\Delta x)/k_BT-\epsilon^2/k_B^2T^2.
\label{eqn:k_mod}
\end{equation}
This relationship suggests that roughness scale $\epsilon$ be extracted if $\log{k(f)}$ is measured over range of temperatures. Variations in temperature also result in changes in the 
viscosity, $\eta$, and 
because $k_0^{-1}\propto\eta$, corrections arising from the temperature-dependence of $\eta$  has to be taken into account in
interpreting the experiments.
It is known that $\eta$ for water varies as $\exp(A/T)$ over the experimentally relevant temperature range ($5^oC<T<50^oC$) \cite{CRC}. 
Thus, we expect $\log{k(f,T)}=a+b/T-\epsilon^2/T^2$. 
The coefficient of the $1/T^2$ term can be quantified by performing \emph{force-clamp experiments} at several values of constant temperatures. 
In addition, the robustness of the HT theory can be confirmed by showing that $\epsilon^2$ is a constant even if the  coefficients $a$ and $b$ change under 
different force conditions \cite{HyeonPNAS03}. The signature of the roughness of the underlying energy landscape is uniquely reflected in the
non-Arrhenius temperature dependence of the unbinding rates. Although it is most straightforward to extract $\epsilon$ using Eq.\ref{eqn:k_mod}
no roughness measurement, to the best of our knowledge, has been performed using force clamp experiments.

To extract the roughness scale, $\epsilon$, using dynamic force spectroscopy (DFS) in which the force increase gradually in time, an alternative but similar 
strategy as in force clamp experiments can be adopted.
A series of dynamic  force spectroscopy experiments should be performed as 
a function of $T$ and $r_f$ so that reliable unfolding force distributions are 
obtained. 
Since a straightforward application of Eq.\ref{eqn:most_force} is difficult due to the force-dependence of the variables in Eq.\ref{eqn:most_force}, one should simplify the expression by assuming that the parameters $\Delta x(f)$, $\Delta F_0^{\ddagger}(f)$, and $\nu_D(f)$, depend only weakly on $f$. 
If this is the case then  the second term of Eq.\ref{eqn:most_force} can be neglected and Eq.\ref{eqn:most_force} becomes 
\begin{equation}
f^*\approx\frac{k_BT}{\Delta x}\log{r_f}+\frac{k_BT}{\Delta x}\log{\frac{\Delta x}{\nu_De^{-\Delta F_0^{\ddagger}/k_BT} k_BT}}+\frac{\epsilon^2}{\Delta x k_BT}.
\label{eqn:most_force_approx}
\end{equation}
One way of obtaining the roughness scale from experimental data is as follows \cite{ReichEMBOrep05}.  From the $f^*$ vs $\log{r_f}$ curves at two different temperatures, $T_1$ and $T_2$, 
one can obtain $r_f(T_1)$ and $r_f(T_2)$ for which the  $f^*$ values are identical. 
By equating the right-hand side of the expression in Eq.\ref{eqn:most_force} at $T_1$ and $T_2$ 
the scale $\epsilon$ can be estimated \cite{HyeonPNAS03,ReichEMBOrep05} as 
\begin{small}
\begin{eqnarray}
\epsilon^2 &\approx&\frac{\Delta x(T_1)k_BT_1\times\Delta x(T_2)k_BT_2}{\Delta x(T_1)k_BT_1-\Delta x(T_2)k_BT_2}\nonumber\\
&\times& \left[\Delta F^{\ddagger}_0\left(\frac{1}{\Delta x(T_1)}-\frac{1}{\Delta x(T_2)}\right)+\frac{k_BT_1}{\Delta x(T_1)}\log{\frac{r_f(T_1)\Delta x(T_1)}{\nu_D(T_1)k_BT_1}}-\frac{k_BT_2}{\Delta x(T_2)}\log{\frac{r_f(T_2)\Delta x(T_2)}{\nu_D(T_2)k_BT_2}}\right]. 
\label{eqn:epsilon}
\end{eqnarray}
\end{small}

In a recent study Nevo et. al. \cite{ReichEMBOrep05} used DFS to measure $\epsilon$ 
for a biomolecular protein complex consisting of nuclear import receptor importin-$\beta$ (imp-$\beta$) and the Ras-like GTPase Ran that is loaded with non-hydrolyzable GTP analogue (Fig.\ref{Nevo_fig}-C).  
The Ran-imp-$\beta$ complex was immobilized on a surface and the unbinding forces were measured using the AFM at three values of $r_f$ that varied by nearly three orders of magnitude.  
At high values of $r_f$ the values of $f^*$ increases as $T$ increases. 
At lower loading rates ($r_f\lesssim 2\times 10^3$ $pN/s$), however, $f^*$ decreases as $T$ increases (see Fig.\ref{Nevo_fig}-B).
The data over distinct temperatures were  used to extract, for the first time, an estimate of $\epsilon$.  The values of $f^*$ at three temperatures (7, 20, 32$^oC$) and Eq.\ref{eqn:epsilon} were used to obtain  $\epsilon\approx 5-6k_BT$.  
Nevo et. al. explicitly showed that the value of $\epsilon$ was nearly the same  from the nine pairs of data extracted from the  $f^*$ vs $\log{r_f}$ curves. 
Interestingly, the estimated value of $\epsilon$ is about $0.2 \Delta F^{\ddagger}_0$ where $\Delta F^{\ddagger}_0$ is the major barrier for unbinding of the complex. 
This shows that, for this complex the free energy in terms of a one-dimensional coordinate, resembles the profile shown in Fig.\ref{landscape}.  
It is worth remarking that the location of the transition state decreases from 0.44 nm at 7$^oC$ to 0.21 nm at 32$^oC$.  
The extracted TS movement using the roughness model is consistent with Hammond behavior (see below).

\section{Extracting TS location ($\Delta x$) and unfolding rate ($\kappa$) using theory of DFS}
The theory of DFS, $f^*\approx\frac{k_BT}{\Delta x}\log{r_f}+\frac{k_BT}{\Delta x}\log{\frac{\Delta x}{\kappa k_BT}}$,  is used to identify the forces that destabilize the bound state of the complex or the folded state of a specific biomolecule.   
A  linear regression  provides the characteristic extension  $\Delta x$ at which the molecule or complex ruptures (more precisely $\Delta x$ is the thermally averaged distance 
between the bound and the transition state along the direction of the applied force). 
It is tempting to obtain the zero force unfolding rate  $\kappa$ from the intercept with the abscissa. 
Substantial errors can, however, arise in the extrapolated values of $\Delta x$ and $\kappa$ to the zero force if $f^*$ vs $\log{r_f}$ is not linear, as is often the case when $r_f$ is varied over several orders of magnitude.  
Nonlinearity of $[f^*,\log{r_f}]$ curve arises for two reasons. One is due to the complicated molecular response to the external load that results in dramatic variations in $\Delta x$.  Like other soft matter,  the extent of response (or their elasticity) depends on $r_f$ \cite{HyeonBJ06,Lacks05BJ,West06BJ}.
The other is due to multiple energy barriers that are encountered in the unfolding or unbinding process \cite{EvansNature99}.

If the TS ensemble is broadly distributed along the reaction coordinate then the molecule can adopt diverse structures along the energy barrier depending on the magnitude of the external load.
Therefore, mechanical force should grasp the signature of the spectrum of the TS conformations for such a molecule. Mechanical unzipping dynamics of RNA hairpins whose stability is determined in terms of the number of intact base pairs is a good example. 
The conformation of RNA hairpins at the barrier top 
can gradually vary from an almost fully intact 
structure at small forces to an extended structure at large forces.  
Under these conditions the width of the TSE is large. 
The signature of diverse TS conformations manifests itself as a substantial 
curvature over the broad variations of forces or loading rates. 
Meanwhile, if the unfolding is a highly cooperative all-or-none process characterized by a narrow distribution of the TS, the nature of the TS may not change significantly.

The linear theory of DFS is not reliable if the TSE is plastic because it involves  drastic approximations of the Eq.\ref{eqn:most_force}. 
From this perspective it is more prudent to
fit the  the experimental unbinding force distributions directly using  analytical expressions derived from suitable models \cite{Dudko06PRL} (see also APPENDIX). 
If such a procedure can be reliably implemented then  
the extracted parameters are likely to be more accurate. 
Solving such an inverse problem does require assuming a reduced dimensional representation of the underlying energy landscape which cannot be \textit{a priori} justified. 

\section{Direct Analysis of the unbinding force distribution}
Although the direct fit of the measured unbinding force distributions to the $P(f)$$\left(=\frac{1}{r_f}k(f)\exp{(-\int^f_0df'\frac{1}{r_f}k(f'))}\right)$ is numerically more complicated than using 
Eq.\ref{eqn:most_force_approx}, it avoids
a potentially serious error due to  the approximation in going from Eq.\ref{eqn:most_force} to Eq.\ref{eqn:most_force_approx}. 
It is, however, difficult to unearth the energy landscape characteristics of the molecule from $P(f)$ alone because $P(f)$ cannot be expressed in a simple functional form. 
By approximating $P(f)$ by a  Gaussian we can dissect the effect of molecular topology and sequence on $P(f)$. 
Note that in general the measured $P(f)$ is asymmetric so that the Gaussian  approximation may be of limited utility. 
For purposes of illustrating the nuances in the data analysis,  we approximate the  measured $P(f)$ by a  Gaussian  using Taylor expansion,
\begin{eqnarray}
P(f)&=&\exp\left[-\left(\int^f_0df\frac{k(f)}{r_f}-\log{\frac{k(f)}{r_f}}\right)\right]\nonumber\\
&=&\exp\left[\frac{\kappa}{r_f}\frac{1}{\beta\Delta x}(e^{\beta f\Delta x}-1)-\beta f\Delta x-\log{\frac{\kappa}{r_f}}\right]\nonumber\\
&\sim&\exp\left[-\frac{(f-\frac{r_f-\kappa k_BT/\Delta x}{\kappa})^2}{2\left(\frac{r_fk_BT}{\kappa\Delta x}\right)}\right]
=\exp{\left(-\frac{(f-\overline{f})^2}{2\sigma_f^2}\right)}
\label{eqn:Pf_Gaussian}
\end{eqnarray}
where $k(f)=\kappa e^{\beta f\Delta x}$ and $\kappa\equiv k_0e^{-\beta \Delta F_0^{\ddagger}}e^{-\beta^2\epsilon^2}$. In going from the first to the second line we assume that $\Delta x$ and $\kappa$ are independent of $f$. 
Two conclusions can be drawn from Eq.\ref{eqn:Pf_Gaussian}.
(i) Both $f^*\sim \overline{f}(=r_f/\kappa-k_BT/\Delta x)$ and the width of the force distribution $\sigma_f(=(\frac{r_fk_BT}{\kappa\Delta x})^{1/2})$ increase as $r_f$ increases. (ii) As $T$ increases, $f^*$ decreases while $\sigma_f$ increases, \emph{provided $\Delta x$ and $\kappa$ are independent function of $T$ or $r_f$}.  

Recently, Schlierf and Rief (SR)  analyzed the unfolding force distributions (with $r_f$ fixed)  of a single domain of \emph{Dictyostelium discoideum} filamin (ddFLN4) at five different temperatures to infer the underlying one dimensional free energy surface \cite{RiefJMB05}.
Going from the measured data ($P(f)$) to the underlying energy landscape is  an inverse problem that requires a specific model.  
Using the Bell model ($k(f)=\kappa\exp{(f\Delta x/k_BT)}$) and an 
approximation that the Arrhenius pre-factor $k_0$ is a constant ($\kappa=k_0\exp{(-\Delta F^{\ddagger}/k_BT)}$ where $k_0=10^7s^{-1}$) that is independent of temperature (i.e., assuming zero roughness), Schlierf and Rief extracted the energy landscape parameters ($\Delta x$, $\Delta F^{\ddagger}$) for each force distribution.  
Somewhat surprisingly, the SR analysis indicated that the rate (with relatively large errors) for populating an intermediate at zero force ($\kappa$) increases as $T$ is lowered.  
Their fits indicate that the location of the TS \textit{must increase} as $T$ increases.  Therefore, they concluded that  ddFLN4 protein exhibits an anti-Hammond behavior, i.e., the position of the TS moves towards the force-stabilized ensemble of states (unfolded states).

Visual inspection of $P(f)$ at different temperatures in Figure 2 of ref. \cite{RiefJMB05} by SR confirms that $f^*$ decreases but $\sigma_f$ decreases as $T$ increases. 
This means that $\kappa\times\Delta x$ must increase faster than the increase of $k_BT$ (Note that $\sigma_f=\left(\frac{r_fk_BT}{\kappa\Delta x}\right)^{1/2}$). 
Although the data were fit well by assuming that 
\emph{$\Delta F^{\ddagger}$ is temperature dependent $\Delta F^{\ddagger}$ while $k_0$ is a constant}, 
an alternative explanation was also possible as recognized by SR.  In the alternate model based on energy landscape roughness we assume that $\Delta F^{\ddagger}$ is constant 
but $k_0$ depends on temperature \cite{HyeonPNAS03}.  The constancy of $\Delta F^{\ddagger}$ is \textit{a posteriori} justified in light of the SR analysis. By adopting the  the HT roughness model SR showed the data can be fit using $\epsilon = 4k_BT$ for ddFLN4 unfolding. 
We believe that SR's main interpretation  (temperature softening effect, or anti-Hammond behavior) still holds even within the framework of the roughness model because their  conclusion is entirely based on  the temperature dependence of $\Delta x$. The roughness model has the advantage that the data can be fit essentially with only one parameter. Moreover, experiments on protein folding have often been interpreted using changes in the prefactor due to $\epsilon$.

To further confirm SR's conclusion we suggest an independent measurement of the precise values of $\Delta x$ by varying not only the temperature but also the loading rate, as is demonstrated by Nevo \emph{et. al.} 
(Fig.\ref{Nevo_fig}-B) \cite{ReichEMBOrep05}. 
The slopes of Fig.\ref{Nevo_fig} at different temperatures succinctly show 
that imp-$\beta$-RanGppNHp complex exhibit the more common  ``Hammond''-behavior. 
Instead of using Eq.\ref{eqn:epsilon}, the energy landscape roughness $\epsilon$ can be extracted from the multiple sets of $P(f)$ at various $(T,r_f)$ conditions. 
Non-Arrhenius behavior characterized particularly by $1/T^2$ is the unique feature of the roughness scale of the energy landscape, and is not affected by TS movements. 
More explicitly, $P(f)$ should be modeled as 
\begin{equation}
P(f)=\frac{\kappa(T;r_f)}{r_f}e^{f\Delta x(T;r_f)/k_BT}\exp{\left[-\int^f_0df'\frac{\kappa(T;r_f)}{r_f}e^{f\Delta x(T;r_f)/k_BT}\right]}
\end{equation}
with $\kappa(T;r_f)=\nu_D(T)e^{-\Delta F^{\ddagger}/k_BT}e^{-\epsilon^2/k_B^2T^2}$. 
The overall shape of $P(f)$ is determined by $\Delta x(T;r_f)$ and $\kappa(T;r_f)$. 
The multiple force distributions, generated at the same loading rate but at 
different temperatures,  can be used to fit $\kappa(T;r_f)$ using 
\begin{equation}
\log{\kappa(T)}=a+b/T-\epsilon^2/T^2.
\end{equation}
Such experiments would permit unambiguous extraction of  roughness just as in the conceptually simpler  force-clamp experiments (compare with Eq.\ref{eqn:k_mod}).

The SR results illustrate the potential difficulties in uniquely extracting model-free energy landscape parameters from dynamic force spectroscopy data. 
It is worth mentioning that Scalley and Baker have used  similar  arguments about the \emph{anomalous 
temperature dependence} of refolding rates of proteins  \cite{Scalley97PNAS}. They showed that the Arrhenius behavior of kinetics is retrieved when the protein stability is corrected for the temperature dependence.

\section{Mechanical response of hard (brittle)  versus soft (plastic) biomolecules}

Regardless of the model used, it is obvious that the lifetimes of a complex decrease upon application of force. The compliance of the molecule is determined by the location of the TS, and hence it is important to understand the characteristics of the molecule that determine the TS.  As we pointed out, many relevant paramters have strong dependence on $f$, $r_f$, or $T$. 
Thus, it is difficult to extract the energy landscape parameters without a suitable model.  In this section we illustrate  two extreme cases of mechanical response \cite{West06BJ,HyeonBJ06,RitortPRL06,West06BJ} of a biomolecule using one-dimensional energy profiles.  In one example the location of the TS does not move with force whereas in the other there is a dramatic movement of the TS. In the presence of force $f$, a given  free energy profile $F_0(x)$ changes to $F(x)$ =  $F_0(x)-fx$.  The location of the TS at non-zero values of $f$ depends on the shape of barrier in the vicinity of the TS.  Near the barrier ($x \approx x_{ts}$) we can approximate $F_0(x)$ as 
\begin{equation}
F_0(x) \approx F_0(x_{ts})-\frac{1}{2} F_0^{\prime\prime}(x_{ts}) (x - x_{ts})^2+\cdots. 
\label{eqn:expansion}
\end{equation}
In the presence of force the TS location becomes $x_{ts}(f)=x_{ts}-\frac{f}{F_0''(x_{ts})}$.  If the transition barrier in $F_0(x)$ is sharp ($x_{ts}F_0''(x_{ts})\gg f$) then we expect very little force-induced movement in the TS.  We refer to molecules that satisfy this criterion as hard or brittle.  In the opposite limit the molecule is expected to be soft or plastic so that there can be dramatic movements in the TS.  We illustrate these two cases by numerically computing $r_f$-dependent  $P(f)$  using Eq.\ref{eqn:MFPT_force} and Eq.\ref{eqn:most_force} for two model free energy profiles. 

{\it Hard response:} A nearly stationary TS position (independent of $f$) 
is realized if the energy barrier is sharp (Eq.\ref{eqn:expansion}). 
We model $F_0(x)$ using 
\begin{equation}
\begin{array}{ll}
     F_0(x)=-V_0|(x+1)^2-\xi^2| & \mbox{with $x\geq 0$}
\end{array}
\label{eqn:hard}
\end{equation}
where $V_0=28$ $pN/nm$ and $\xi=4$ $nm$. 
The energy barrier forms at $x=1$ $nm$ and this position does not change much even in the presence of force as illustrated in Fig.\ref{hardplot.fig}-A.  In dynamic force spectroscopy the free energy profiles drawn at constant force may be viewed as  snapshots at different times. 
The shape of the unbinding force distribution  depends on $r_f$. 
We calculated $P(f)$  numerically using Eq.\ref{eqn:MFPT_force} and Eq.\ref{eqn:most_force} (see Fig.\ref{hardplot.fig}-B). 
Interestingly, a plot of the the most probable force $f^*$ obtained from $P(f)$ does not exhibit any curvature when $r_f$ is varied over
six orders of magnitude (Fig.\ref{hardplot.fig}-C). 
Over the range of $r_f$ the $[f^*, \log{r_f}]$ plot is almost linear. 
The slight deviation from linearity is due to the force-dependent curvature near the bound state ($\omega_b(f)$). 
From the slope we find that  $\Delta x\approx 1$ $nm$ which is expected from Eq.\ref{eqn:hard}. 
In addition,  we obtained from the intercept in Fig.\ref{hardplot.fig}-C that $\kappa=1.58$ $s^{-1}$.  
The value of $\kappa[\equiv k(f=0)]$ directly computed using 
Eq.\ref{eqn:MFPT_force} is $\kappa=1.49$ $s^{-1}$. 
The two values agree quite well. 
Thus, for brittle response the Bell model is expected to be accurate.  

{\it Soft response:} If the  position of the TS \emph{sensitively} moves with force the biomolecule or the complex is soft or plastic. 
To illustrate the behavior of soft molecules we model the free energy potential in the absence of force using 
\begin{equation}
\begin{array}{ll}
     F_0(x)=-V_0\exp{(-\xi x)} & \mbox{with $x\geq 0$}
\end{array}
\label{eqn:soft}
\end{equation}
where $V_0=82.8$ $pN\cdot nm$ and $\xi=4$ $(nm)^{-1}$. 
The numerically computed $P(f)$ and $[f^*, \log{r_f}]$ plots are shown Fig.\ref{softplot.fig}.  The slope of the $[f^*, \log{r_f}]$ plot is no longer constant but increases continuously as $r_f$ increases. 
The extrapolated value of $\kappa$ to zero $f$ varies greatly depending on the range of $r_f$ used.  Even in the experimentally accessible range of $r_f$ there is curvature in the $[f^*,\log{r_f}]$ plot.  Thus, unlike the parameters ($\Delta x$, $\kappa$) in the example of a brittle potential, all the extracted parameters from the force profile  are strongly dependent on the loading rate. 
As a result, in soft molecules the extrapolation to zero force (or minimum loading rate) is not as meaningful as in hard molecules. 
Note how the extracted $\Delta x$ (see the inset of Fig.\ref{softplot.fig}-A) changes as a function of $r_f$. 
For soft (plastic) molecules, the extracted parameters using the tangent at a certain $r_f^o$ are not the characteristics of the free energy profile in the absence of the load, but reflect the  features for the modified free energy profile tilted by $(f^*)^o$ at $r_f^o$.

In practice, biomolecular systems lie between the two extreme cases (brittle and plastic).  
In many cases the $[f^*, \log{r_f}]$ appears to be linear over a narrow range of $r_f$. 
The linearity in narrow range of $\log{r_f}$, however, does not guarantee the linearity under broad variations of loading rates.  
In order to obtain energy landscape  parameters it is important to perform experiments 
at $r_f$ as low as possible.  
The brittle nature of proteins (lack of change in $\Delta x$) inferred from AFM experiments may be the result of a relatively large $r_f$ ($\approx 1,000 pN/s$).  
On the other hand, only by varying $r_f$ over a wide range the molecular elasticity of proteins and RNA 
can be completely described.  
Indeed, we showed that even in simple RNA hairpins the transition 
from plastic to brittle behavior can be achieved by varying $r_f$ \cite{HyeonBJ06}. 
The load-dependent response may even have  functional significance.

\section{Hammond/anti-Hammond behavior under force and temperature variations}
The qualitative nature of the TS movement with increasing perturbations can often be anticipated using the Hammond postulate which has been successful in not only analyzing a large class of chemical reactions but also in rationalizing the observed behavior in protein and RNA foldings.  
The Hammond postulate states that the nature of TS resembles the least stable species along the reaction pathway. In the context of forced-unfolding it  implies that the TS location should move closer to the native state as $f$ increases.  
In other words $\Delta x$ should decrease as $f$ is increased.  Originally Hammond postulate was introduced to explain chemical reactions involving  small organic molecules \cite{HammondJACS53,LefflerSCI53}.  
Its validity in biomolecular folding is not obvious because  there are  multiple folding or unfolding pathways.  
As a result there is a large entropic component to the folding reaction.  Surprisingly, many folding processes are apparently in accord with the Hammond postulate \cite{Fersht95Biochemistry,Dalby98Biochemistry,Kiefhaber00PNAS}. If the extension is an appropriate reaction coordinate for forced unfolding then deviations from Hammond postulate should be an exception than the rule.  Indeed, anti-hammond behavior (movement of the TS closer to more stable unfolded state as $T$ increases) was suggested by SR based on a model used to analyze the AFM data.  
The simple free energy profiles used in the previous section (Eq.\ref{eqn:hard} and Eq.\ref{eqn:soft}) can be used to verify the Hammond postulate when the external perturbation is either force or temperature. 
First, for the case of hard response the TS is barely affected by force, thus the Hammond or anti-Hammond behavior is not a relevant issue when unbinding is induced by $f$.  
On the other hand, for the case of soft molecules $\Delta x$ always decreases with a larger force. 
The positive curvature in $[f^*,\log{r_f}]$ plot is the signature of the classical Hammond-behavior with respect to $f$. 

As long as a one dimensional free energy profile suffices in describing forced-unfolding of proteins and RNA the TS location must satisfy Hammond postulate.  In general, for a fixed force or $r_f$, $\Delta x$ can vary with $T$. The changes in $\Delta x$ with temperature can be modeled using $T$-dependent parameters in the potential. To evaluate the consequence of $T$-variations we set  
\begin{equation}
\xi=\xi_0+\alpha(T-300K)
\end{equation} 
for both free energies in Eq.\ref{eqn:hard} and Eq.\ref{eqn:soft}. 
Depending on the value of $\alpha$  the position of the TS can move towards or away from the native state. We set $\alpha=\pm 0.1$ for both the hard and soft cases. The numerically computed   $[f^*, \log{r_f}]$ plots are shown in Fig.\ref{hard_soft_Hammond.fig}. 
One interesting point is found in soft molecule that exhibits Hammond behavior.  For wide range of $r_f$, $\Delta x$ decreases as $T$ increases. However, the most probable unbinding force $f^*$ at low temperatures can be larger or smaller than $f^*$ at high temperatures depending on the loading rate (see upper-right corner of Fig.\ref{hard_soft_Hammond.fig}).  A very similar behavior has been observed in the forced-unbinding of Ran-imp-$\beta$ complex \cite{ReichEMBOrep05} (see also Fig.\ref{Nevo_fig}-B).
Although the model free energy profiles can produce a wide range of behavior depending on $T$, $f$, and $r_f$ the challenge is to provide a structural basis for the measurements on biomolecules.  

\section{Multidimensionality of energy landscape coupled to "memory" affect the force dynamics}
The natural one dimensional reaction coordinate in  mechanical unfolding experiments is the extension $x$ of the molecule.  
However, local rupture events can couple to $x$, which would require a multidimensional description. 
For example, consider the case of forced-unfolding of a nucleic acid hairpin.  
The opening of a given base pair is dependent not only on its strength, which resists unfolding, but also on the increase in the available conformational space which favors unfolding.  
In this case fluctuations in the collective coordinates, which describe the local events, 
are coupled to the global coordinate $x$.  
Thus, suitable fluctuations in the local coordinate $r$ has to occur before an increase in the extension is observed.  
Such a coupling between local coordinates and global observable arises naturally in many physical situations \cite{Zwanzig90ACR}. 
In the context of ligand binding to myoglobin Zwanzig first proposed such a model by assuming 
that  reaction (binding) takes place along $x$-coordinate and at the barrier top ($x=x_{ts}$) the reactivity is determined by the cross section of bottleneck described by $r$-coordinate \cite{Zwanzig92JCP}. 
We adopt a similar picture to describe the modifications when such a process is driven by force. 
First, consider the Zwanzig case i.e, $r_f = 0$.
The equations of motion for $x$ and $r$ are, respectively, 
\begin{eqnarray}
m\frac{d^2x}{dt^2}&=&-\zeta\frac{dx}{dt}-\frac{dU(x)}{dx}+F_x(t)\nonumber\\
\frac{dr}{dt}&=&-\gamma r+F_r(t).
\label{eqn:eqofmotion}
\end{eqnarray}
The Liouville theorem ($\frac{d\rho}{dt}=0$) describes the time evolution of probability density, $\rho(x,r,t)$, as 
\begin{equation}
\frac{d\rho}{dt}=\frac{\partial\rho}{\partial t}+\frac{\partial}{\partial x}\left(\frac{dx}{dt}\rho\right)+\frac{\partial}{\partial r}\left(\frac{dr}{dt}\rho\right)=0.
\label{eqn:Liouville}
\end{equation}
 By inserting of Eq.\ref{eqn:eqofmotion} to Eq.\ref{eqn:Liouville} and  neglecting the inertial term, ($m\frac{d^2x}{dt^2}$), 
and averaging over the white-noise spectrum, 
and the fluctuation-dissipation theorem
($\langle F_x(t)F_x(t')\rangle=2\zeta k_BT\delta(t-t')$, $\langle F_r(t)F_r(t')\rangle=2\lambda\theta\delta(t-t')$ 
where $\langle r^2\rangle\equiv \theta$) leads to a Smoluchowski equation 
for $\rho(x,r,t)$ in the presence of a reaction sink. 
\begin{equation}
\frac{\partial\overline{\rho}}{\partial t}=\mathcal{L}_x\overline{\rho}+\mathcal{L}_r\overline{\rho}-k_rr^2\delta(x-x_{ts})\overline{\rho}
\label{eqn:Smol}
\end{equation} 
where $\mathcal{L}_x\equiv D\frac{\partial}{\partial x}\left(\frac{\partial}{\partial x}+\frac{1}{k_BT}\frac{dU(x)}{dx}\right)$ and $\mathcal{L}_r\equiv \lambda\theta\frac{\partial}{\partial r}\left(\frac{\partial}{\partial r}+\frac{r}{\theta}\right)$. 
Integrating  both sides of Eq.\ref{eqn:Smol} using $\int dx\rho(x,r,t)\equiv \overline{C}(r,t)$ leads to 
\begin{equation}
\frac{\partial\overline{C}}{\partial t}=\mathcal{L}_r\overline{C}(r,t)-k_rr^2\overline{\rho}(x_{ts},r,t). 
\label{eqn:step1}
\end{equation}
By writing 
$\overline{\rho}(x_{ts},r,t)=\phi_x(x_{ts})\overline{C}(r,t)$ 
where $\phi(x_{ts})$ should be constant as 
$\phi(x_{ts})=e^{-U(x_{ts})/k_BT}/\int dx e^{-U(x)/k_BT}\approx \sqrt{\frac{U''(x_b)}{2\pi k_BT}}e^{-(U(x_{ts})-U(x_b))/k_BT}$, 
 Eq.\ref{eqn:step1} becomes
\begin{equation}
\frac{\partial\overline{C}}{\partial t}=\mathcal{L}_r\overline{C}(r,t)-kr^2\overline{C}(r,t). 
\label{eqn:step2}
\end{equation}
where $k\equiv k_r\sqrt{\frac{U''(x_b)}{2\pi k_BT}}e^{-(U(x_{ts})-U(x_b))/k_BT}$. In all likelihood $k_r$ reflects the dynamics near the barrier, 
so we can write $k=\kappa \frac{\omega_{ts}\omega_b}{2\pi\gamma}e^{-\Delta U/k_BT}$ where $\kappa$ describes the geometrical information of the cross section of bottleneck. Now we have retrieved the equation in Zwanzig's seminal paper where the 
survival probability ($\Sigma(t)=\int^{\infty}_0 dr\overline{C}(r,t)$) is given under a reflecting boundary condition at $r=0$ and Gaussian initial condition $\overline{C}(r,t=0)\sim e^{-r^2/2\theta}$. 
By setting $\overline{C}(r,t)=\exp{(\nu(t)-\mu(t)r^2)}$, Eq.\ref{eqn:step2} can be solved exactly, leading to 
\begin{eqnarray}
\nu'(t)&=&-2\lambda\theta\mu(t)+\lambda\nonumber\\
\mu'(t)&=&-4\lambda\theta\mu^2(t)+2\lambda\mu(t)+k.
\end{eqnarray}
The solution for $\mu(t)$ is obtained by solving 
$\frac{4\theta}{\lambda}\int^{\mu(t)-1/4\theta}_{1/4\theta}\frac{d\alpha}{\sigma^2-16\theta^2\alpha^2}=t$, and this leads to 
\begin{eqnarray}
\frac{\mu(t)}{\mu(0)}&=&\frac{1}{2}\left\{1+S\frac{(S+1)-(S-1)E}{(S+1)+(S-1)E}\right\}\nonumber\\
\nu(t)&=&-\frac{\lambda t}{2}(S-1)+\log{\left(\frac{(S+1)+(S-1)E}{2S}\right)^{-1/2}}
\end{eqnarray}
with $\mu(0)=1/2\theta$.
The survival probability, which was derived by Zwanzig,  is 
\begin{equation}
\Sigma (t)=\exp{\left(-\frac{\lambda}{2}(S-1)t\right)}
\left(\frac{(S+1)^2-(S-1)^2E}{4S}\right)^{-1/2}
\label{eqn:solution}
\end{equation}
where $S=\left(1+\frac{4k\theta}{\lambda}\right)^{1/2}$ and $E=e^{-2\lambda St}$.

We wish to examine the consequences of coupling between local and global reaction coordinates under tension. In order to accomplish our goal we solve the Smoluchoski equation in the presence of constant load. In this case, 
$k$ in Eq.\ref{eqn:step2} should be replaced with $ke^{t (r_f\Delta x/k_BT)}$. 
Eq.\ref{eqn:step2}i, however, becomes hard to solve if the sink term depends on $t$. 
Nevertheless, analytical solutions can be obtained for special cases of $\lambda$. 
If $\lambda\rightarrow \infty$, $d\Sigma(t)/dt=-k\theta e^{t(r_f\Delta x/k_BT)}\Sigma(t)$, and hence, 
\begin{equation}
\Sigma(f)=\exp{\left(-k\theta\int^f_0df\frac{1}{r_f}e^{f\Delta x/k_BT}\right)}=\exp{\left(-\frac{k\theta k_BT}{r_f\Delta x} (e^{f\Delta x/k_BT}-1)\right)}
\end{equation}
Using the  rupture force distribution $P(f)=-\frac{d\Sigma(f)}{df}$ and $\frac{dP(f)}{df}|_{f=f^*}=0$, one can obtain 
the most probable force 
\begin{equation}
f^*=\frac{k_BT}{\Delta x}\log{\frac{r_f\Delta x}{(k\theta)k_BT}}.
\end{equation}

If $\lambda$ is small ($\lambda\rightarrow 0$) then 
$\overline{C}(r,t)=\exp{[-\int_0^t dt kr^2\exp{(t\times r_f\Delta/k_BT)}]}=\exp{\left[-\frac{kr^2 k_BT}{r_f\Delta x}(e^{tr_f\Delta x/k_BT}-1)\right]}$ with the initial distribution of $e^{-r^2/2\theta}$.
Thus,
\begin{eqnarray}
\Sigma(f)&=& \int_0^{\infty}dr\exp{\left[-r^2\frac{kk_BT}{r_f\Delta x}(e^{f\Delta x/k_BT}-1)\right]}\sqrt{\frac{2}{\pi\theta}}\exp{\left[-r^2/2\theta\right]}\nonumber\\
&=& \sqrt{\frac{2}{\pi\theta}}\left(\frac{kk_BT}{r_f\Delta x}(e^{f\Delta x/k_BT}-1)+\frac{1}{2\theta}\right)^{-1/2}. 
\end{eqnarray}
Note that if $r_f\rightarrow 0$ we recover Zwanzig's result $\Sigma(t)\sim (1+2k\theta t)^{-1/2}$. Using $P(f)=-d\Sigma(f)/df$
\begin{equation}
P(f)=\frac{1}{\sqrt{2\pi\theta}}\left(\frac{kk_BT}{r_f\Delta x}(e^{f\Delta x/k_BT}-1)+\frac{1}{2\theta}\right)^{-3/2}\frac{k}{r_f}e^{f\Delta x/k_BT}. 
\end{equation}
$dP(f)/df|_{f=f^*}=0$ gives 
\begin{equation}
f^*=\frac{k_BT}{\Delta x}\log{\left\{\left(\frac{r_f\Delta x}{k\theta k_BT}\right)\left(1-\frac{2k\theta k_BT}{r_f\Delta x}\right)\right\}}, 
\label{eqn:smalllambda}
\end{equation}
in which $r_f\geq(1+2\theta)\frac{k\theta k_BT}{\Delta x}$ since $f^*\geq 0$.
This shows that $f^*$ vs $r_f$ has a different form when $\lambda\rightarrow 0$ from the one when $\lambda\rightarrow\infty$.
The deviation of Eq.\ref{eqn:smalllambda} from the conventional relation is pronounced when $\left[r_f-(1+2\theta)\frac{kk_BT}{\Delta x}\right]\rightarrow 0^+$.

At present, experimental data have been interpreted using only one dimensional free energy profiles.  
The meaning and the validity of the extracted free energy profiles has not been established. 
At the least, this would require computing the force-dependent first passage times using the "experimental" 
free energy profile assuming that the extension is the only slowly relaxing variable. 
If the computed force-dependent rates (inversely proportional to first passage times) agree with the measured rates then the use of extension of the reaction coordinate would be justified.  
In the absence of good agreement with experiments other models, such as the one we have proposed here, must be considered.  
In the context of force-quench refolding we have shown (see below) that extension alone is not an adequate reaction coordinate.   
For refolding upon force-quench of RNA hairpins, the coupling between extension and local dihedral angles, which reports on the conformation of the RNA, needs to be taken into account to quantitatively describe the refolding rates.    

\section{Conclusions}
With the advent of single molecule experiments that can manipulate biomolecules using mechanical force it has become possible to get a detailed picture of their energy landscapes. Mechanical folding and unfolding trajectories of proteins and RNA show that there is great diversity in the explored routes \cite{HyeonSTRUCTURE06,Bustamante03Science,FernandezSCI04}.   In certain well defined systems with simple native states, such as RNA and DNA hairpins, it has been shown using constant force unfolding that the hairpins undergo sharp bistable transitions from folded to unfolded states \cite{Bustamante02Science,Block06Science,Woodside06PNAS}.  
From the dynamics of the extension as a function of time measured over a long period the underlying force dependent profiles have been inferred.  
The force-dependent folding and unfolding rates and the unfolding trajectories can be used to construct the one-dimensional energy landscape. In a remarkable paper \cite{Block06Science}, Block and coworkers have shown that the location of the TS can be moved, at will, by varying the hairpin sequence.   
The TS was obtained using the Bell model by assuming that the $\Delta x$ is independent of $f$.  
While this seems reasonable given the sharpness of the inferred free energy profiles near the barrier top it will be necessary to show the $\Delta x$ does not depend on force.

The fundamental assumption in inverting the force-clamp data is that the molecular extension is a suitable reaction coordinate.  This may indeed be the case for force-spectroscopy in which the  response of the molecule only depends on force that is coupled to the molecular extension, which may well  
represent the slow degrees of freedom.  
The approximation is more reasonable for forced-unbinding. 
It is less clear if it can be assumed that extension $x$ is the appropriate reaction coordinate when refolding is initiated by quenching the force to low enough values such that the folded state is preferentially populated.  
In this case the dynamic reduction in $x$  can be coupled to collective internal degrees of freedom. 
In a recent paper \cite{HyeonBJ06} we showed, in the context of force-quench refolding of a RNA hairpin, that the reduction in $x$ is largely determined by local conformational changes in the dihedral angle degrees in the loop region.  Zipping by nucleation of the hairpin with concomitant reduction in $x$ does not occur until the transitions from \textit{trans} to \textit{gauche} state in a few of the loop dihedral angles take  place.  
In this case, one has to consider at least a two dimensional free energy landscape. 
Fig.\ref{dih_R_2D_map_illust.fig} clearly shows such a coupling between end-to-end distance ($R$) and the dihedral angle degrees of freedom. The "correctness" of the six dihedral angles representing the conformation of RNA hairpin loop region ($\phi_i$, $i=19,\ldots 24$) is quantified using $\langle 1-\cos{(\phi_i-\phi_i^o)}\rangle$, where $\phi^o_i$ is the angle value in the native state and $\langle\ldots\rangle$ is the average over the six dihedral angles. 
$\langle 1-\cos{(\phi-\phi^o)}\rangle=0$ signifies the correct dihedral conformation for the hairpin loop region. 
Once the "correct" conformation is attained in the loop region, the rest of the zipping process can easily proceed as we have shown in \cite{HyeonBJ06}. 
Before the correct loop conformation being attained, RNA spends substantial time in searching the conformational space related to the dihedral angle degree of freedom. 
The energy landscape ruggedness is manifested as in Fig.\ref{dih_R_2D_map_illust.fig} when conformational space is represented using multidimensional order parameters.   The proposed coupling between the local dihedral angle degrees of freedom and extension (global parameter) is fairly general.  A similar structural slowing down, due to the cooperative link between local and global coordinates, should be observed in force-quench refolding of proteins as well.

The most exciting use of singe molecule experiments is in their ability to extract precise values of the energy landscape roughness $\epsilon$ by using temperature as a variable in addition to $f$.  
In this case a straightforward measurement of the unbinding rates as a function of $f$ or $r_f$ can be used to obtain $\epsilon$ without having to make \textit{any assumptions} about the underlying mechanisms of unbinding.  
Of course, this involves performing a number of experiments. 
In doing so one can also be rewarded with diagram of states in terms of $f$ and $T$ \cite{HyeonPNAS05}.  
The theoretical calculations and arguments given here also show that the power of single molecule experiments can be fully realized only by using  the data in conjunction with  carefully designed theoretical and computational models.  
The latter can provide the structures that are  sampled in the process of forced-unfolding and force-quench refolding as was illustrated for ribozymes and GFP \cite{HyeonSTRUCTURE06}.  
It is likely that the promise of measuring the energy landscapes of biomolecules, almost one molecule at a time, will be fully realized using a combination of single molecule measurements, theory, and simulations.  Recent studies have already given us a glimpse of that promise with more to come shortly.

\section{Appendix: Dynamic Force spectroscopy in a Cubic potential}
Assuming that the Bell model gives a correct description of forced-unbinding it was shown by Evans and Ritchie that the most probable unbinding force $f^* \approx \frac{k_BT}{\Delta x_{ts}}\log{r_f}$ assuming that the TS location is independent of $r_f$ \cite{Evans97BJ}.  
Deviations from logarithmic dependence of $f^*$ on $r_f$ occurs if the assumptions of the Bell model is relaxed as was first shown by Dudko et. al. \cite{Dudko03PNAS}. 
In this Appendix we give a simple derivation of the result  for the cubic potential, the simplest one that  has a potential well (bound state)  and a barrier for which we can  analytically formulate the dynamic force spectroscopy theory. The general form of the cubic potential is,
 \begin{equation}
\begin{array}{ll}
     F_0(x)=ax^3+bx^2+cx+d & \mbox{with $a<0$}.
\end{array}
\end{equation}
In the presence of constant force $f$, 
$F(x)=F_0(x)-fx$ can have a finite free energy barrier if the two roots of $F'(x)= 3ax^2+2bx+c-f=0$, namely, 
\begin{equation}
r_{\pm}=\frac{-2b\pm\sqrt{4b^2-12a(c-f)}}{6a}\nonumber\\
\end{equation}
are real.
Provided $b^2-3a(c-f)>0$, $r_-$ and $r_+$ correspond to $x_{ts}$ and $x_b$, respectively. Note that both $x_{ts}$ and $x_b$ are functions of $f$ in the cubic potential and so is $\Delta x=x_{ts}-x_b=-\frac{\sqrt{4b^2-12a(c-f)}}{3a}$. 
The distance between $x_{ts}$ and $x_b$ decreases as $f$ grows and 
vanishes when the force reaches a critical force $f_c=c-\frac{b^2}{3a}$, where the free energy barrier also becomes zero. 
The $f$-dependent free energy barrier($\Delta F^{\ddagger}$) is calculated by $\Delta F^{\ddagger}(f)=F(x_{ts})-F(x_b)$ as,
\begin{eqnarray}
\Delta F^{\ddagger}(f)&=&a(r_-^3-r_+^3)+b(r_-^2-r_+^2)+c(r_--r_+)-f(r_--r_+)\nonumber\\
&=&\frac{2\sqrt{-12a}}{(-9a)}f_c^{3/2}\left(1-\frac{f}{f_c}\right)^{3/2}\nonumber\\
&=&U_c\epsilon_c^{3/2}.
\label{eqn:cubic_barrier} 
\end{eqnarray}
where $\epsilon\equiv 1-f/f_c$.
The curvatures at the barrier ($x=x_{ts}$) and the bottom ($x=x_b$) of the potential,  
that are used in  the Kramers' rate expression, are calculated 
using $F''(x)|_{x=x_{ts}, x_b}=(6ax+2b)|_{x=x_{ts}, x_b}=m\omega^2_{ts,b}$.
\begin{eqnarray}
\sqrt{m}\omega_{ts}=\sqrt{m}\omega_b&=&\left[-12af_c\left(1-\frac{f}{f_c}\right)\right]^{1/4}\nonumber\\
&=&\Omega_c\epsilon^{1/4}
\label{eqn:cubic_omega}
\end{eqnarray} 
Note that (i) $\Delta x=m\omega^2/(-3a)$ is satisfied for all $0<f<f_c$, and (ii) $\omega_{ts}=\omega_b$, the \emph{point symmetry} around the inflection point $x_c$ that satisfies 
$F'(x_c)=F''(x_c)=0$, 
are the unique properties of  the cubic function for
any parameter set $(a,b,c,d,f)$.  

This property is no longer valid if the free energy function is modeled using a higher order polynomial or special functions. What is worse for the higher order polynomial is that 
roots of $F'(x)=0$ are not easily found, and that if order is higher than 6 there is no general solution. Thus, if higher polynomials are needed to fit the free energy profile, even with the assumption that the extension is an appropriate reaction coordinate, then analytically tractable solutions are not possible. It may be argued that any smooth potential may be locally approximated using a cubic potential, and hence can be used to analyze the experimental data. 

Using Eqs.\ref{eqn:Kramers}, \ref{eqn:force_distribution}, \ref{eqn:cubic_barrier}, and \ref{eqn:cubic_omega} we get
\begin{eqnarray}
k_{cubic}(\epsilon)=\frac{\Omega_c^2\epsilon^{1/2}}{2\pi\gamma}\exp{\left(-U_c\epsilon^{3/2}/k_BT\right)}
\end{eqnarray}
and 
\begin{eqnarray}
P(\epsilon)&=&\frac{1}{r_f}\frac{\Omega_c^2\epsilon^{1/2}}{2\pi\gamma}\exp{\left(-U_c\epsilon^{3/2}/k_BT\right)}\exp{\left(\frac{f_c}{r_f}\frac{\Omega_c^2}{2\pi\gamma}\int^{\epsilon}_1d\epsilon\epsilon^{1/2}e^{-U_c\epsilon^{3/2}/k_BT}\right)}\nonumber\\
&=&\mathcal{N}\epsilon^{1/2}\exp{\left(-\frac{U_c\epsilon^{3/2}}{k_BT}-\frac{k_BT f_c\Omega_c^2}{3\pi U_c\gamma r_f}e^{-U_c\epsilon^{3/2}/k_BT}\right)}
\end{eqnarray}
where $\mathcal{N}=\frac{\Omega_c^2}{2\pi\gamma r_f}\exp{\left(\frac{k_BT f_c\Omega_c^2}{3\pi U_c \gamma r_f}e^{-U_c/k_BT}\right)}$. 
The most probable force $f^*$ is obtained using $dP(f)/df|_{f=f^*}=0$, which 
establishes the relation with respect to $\epsilon^*(\equiv 1-f^*/f_c)$ as 
\begin{equation}
-\frac{U_c}{k_BT}(\epsilon^*)^{3/2}=\log{\left[\frac{\gamma}{\Omega^2_cf_c}\left(\frac{U_c}{k_BT}-(\epsilon^*)^{-2/3}\right)\times r_f\right]}.
\end{equation}
If $f^*\ll f_c$ which also satisfies the condition $\frac{U_c}{k_BT}\gg(\epsilon^*)^{-2/3}$, 
\begin{equation}
f^*\approx f_c\left\{ 1-\left[\frac{k_BT}{U_c}\log{\left(\frac{\gamma}{\Omega^2_cf_c}\frac{U_c}{k_BT}\times r_f\right)}\right]^{2/3}\right\}. 
\end{equation}
By defining $r_f^{min}\equiv \frac{U_c\Omega_c^2f_c}{k_BT\gamma}e^{U_c/k_BT}$ as the minimum loading rate that gives $f^*>0$, one can rewrite the above expression as 
\begin{equation}
f^*\approx f_c\left[1-\left(1+\frac{k_BT}{U_c}\log{\frac{r_f}{r_f^{min}}}\right)^{2/3}\right]. 
\label{eqn:f*approx}
\end{equation}
The importance of the result in Eq.\ref{eqn:f*approx} was first emphasized by Dudko et. al.

Two comments about the announced deviation from the usual logarithmic dependence of $f^*$ on $r_f$ are worth making. (1) Provided $\frac{k_BT}{U_c}\log{\frac{r_f}{r_f^{min}}}\ll 1$, Eq.\ref{eqn:f*approx} becomes
\begin{equation}
f^*\approx \frac{k_BT}{\left(\frac{3U_c}{2f_c}\right)}\log{\frac{r_f}{r_f^{min}}}, 
\end{equation}
which is the typical $f^*$ vs $\log{r_f}$ relation. 
It is of particular interest to see if the condition $\frac{k_BT}{U_c}\log{\frac{r_f}{r_f^{min}}}\ll 1$ is satisfied in proteins or RNA. (2) Because $f^* \approx {log r_f}^{\nu}$ in both the Evans-Ritchie ($\nu = 1$) \cite{Evans97BJ} and Dudko ($\nu = \frac{2}{3}$) formulations \cite{Dudko06PRL} it is unclear whether they can be really distinguished using experimental data in which $r_f$ cannot be easily varied by more than three (at best) orders of magnitude. Thus, the reliability of the parameters obtained using either formulation is difficult to assess independently.  Nevertheless, the reexamination of the logarithmic dependence of $f^*$ on $r_f$ shows that one should be mindful of the assumption that the location of the TS \textit{does not depend on $f$ or $r_f$}. \\

{\bf Acknowledgments:} We are grateful to Reinat Nevo and Ziv Reich for providing the data in Fig. (3).  This work was supported in part by a grant from the National Science Foundation through grant number CHE 05-14056.
\clearpage 
\section*{\bf Figure Caption}
\begin{figure}[ht]
\includegraphics[width=5.00in]{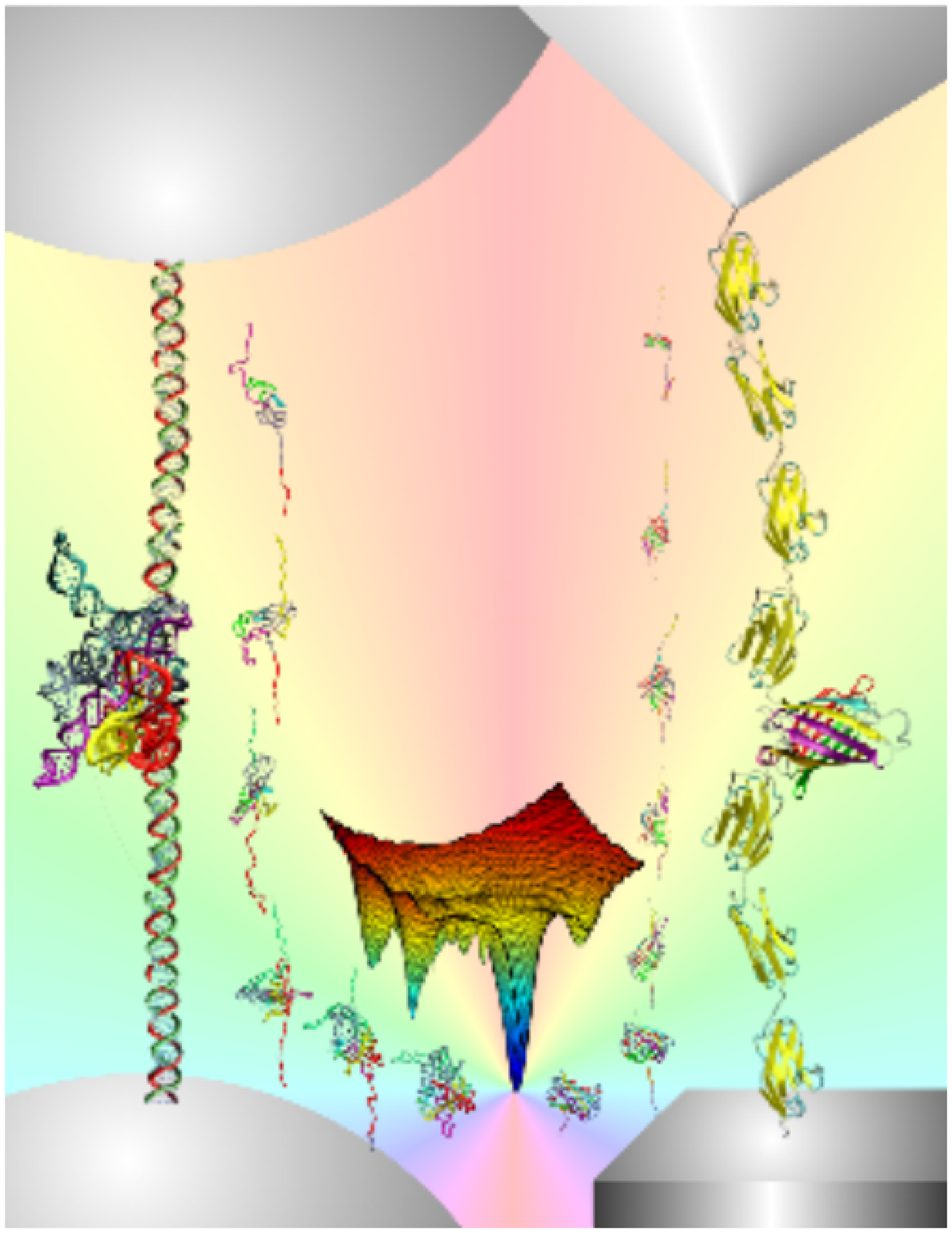}
\caption{\label{RoughreviewFig1.fig}}
\end{figure}

{\bf Figure \ref{RoughreviewFig1.fig} :}
Single molecule experiments using mechanical force. On the left we show a schematic setup of LOT experiments in which a ribozyme is mechanically unfolded.  The \textit{Tetrahymena} ribozyme is sandwiched between two RNA/DNA hybrid handles that are linked to micron-sized beads.  The structures of a few of the displayed intermediates were obtained using simulations of the Self-Organized Polymer (SOP) model representation \cite{HyeonSTRUCTURE06} of the ribozyme.  The sketch on the right shows a typical AFM setup in which we illustrate unfolding of GFP.  The intermediate structures, that are shown, were obtained using simulations of the SOP model for GFP.

{\bf Figure \ref{landscape} :}
Caricature of the rough energy landscape of proteins and RNA that fold in an apparent ''two-state'' manner using extension $x$, the coordinate that is
conjugate to force $f$. 
Under force $f$, the zero-force free energy profile ($F_0(x)$) is tilted 
by  $f\times x$ and gives rise to the free energy profile, $F(x)$. 
In order to clarify the derivation of Eq.\ref{eqn:mfpt} we have explicitly indicated the average location of the relevant parameters.

{\bf Figure \ref{Nevo_fig} :}
Dynamic force spectroscopy measurements of single imp-$\beta$-RanGppNHp pairs at different temperatures. 
{\bf A.} Distributions of measured unbinding forces using AFM for the lower-strength 
conformation of the complex at different loading rates at 7 and 32$^oC$. 
Roughness acts to increase the separation between the distributions recorded at different temperatures. The histograms are fit using Gaussian distributions. 
The width of the bins represents the thermal noise of the cantilever. 
{\bf B.} Force spectra used in the analysis. The most probable unbinding forces $f^*$ are plotted as a function of $log(r_f)$. 
The maximal error is $\pm10$\% because of uncertainities in 
determining the spring constant of the cantilevers. 
Statistical significance of the differences between the slopes of the spectra was confirmed using covariance test. 
(Images courtesy of Reinat Nevo and Ziv Reich \cite{ReichEMBOrep05}). {\bf C.} Ran-importin$\beta$ complex crystal structures (PDB id: 1IBR \cite{Vetter99Cell}) in surface (left) and ribbon (right) representations. In AFM experiments, 
Ran (red) protein complexed to importin$\beta$ (yellow) is pulled until the dissociation of the complex takes place. 

{\bf Figure \ref{hardplot.fig} :}
Dynamic force spectroscopy analysis using a model free energy profile
$F_0(x)=-V_0|(x+1)^2-\xi^2|$ with $V_0=20pN/nm$, $\xi=4nm$, and $x\geq 0$. The lack of change in $x_{ts}$ as $f$ changes shows a \emph{hard response} under tension.  
{\bf A.} Effective free energy profile ($F(x)$) at various values of $f$. 
{\bf B.} Distributions of unbinding forces at different loading rates. 
{\bf C.} Plot of the most probable unbinding force ($f^*$) versus $\log{r_f}$.

{\bf Figure \ref{softplot.fig} :}
Dynamic force spectroscopy for soft response to $f$ using the $f = 0$ free energy profile  
$F_0(x)=-V_0\exp{(-\xi x)}$ with $V_0=82.8pN\cdot nm$, $\xi=4(nm)^{-1}$.   
{\bf A.} Effective free energy profile ($F(x)$) as a function of $f$. 
For emphasis on the soft response of the potential, 
the position of TS at each force value is indicated with arrows. 
{\bf B.} Distributions of unbinding forces at varying loading rates. 
{\bf C.} Plot of most probable unbinding force ($f^*$) versus $\log{r_f}$. The slope of the tangent at each loading rate value varies substantially, which suggests the variation in the TS (inset) as $r_f$ changes.

{\bf Figure \ref{hard_soft_Hammond.fig} :}
By tuning the value of $\xi$ (see Eq. (23)) as a function of temperature, 
Hammond and anti-Hammond behaviors emerge  
in the context of force spectra in the free energy profiles that show hard and soft responses. The condition for Hammond or anti-Hammond
behavior depends on $\alpha$ (Eq. (23)).

{\bf Figure \ref{dih_R_2D_map_illust.fig} :}
{\bf A.} A sample refolding trajectory of a RNA hairpin  starting from the stretched state.
The hairpin was, at first, mechanically unfolded to a fully stretched state and the force was subsequently quenched
 to zero at $t\approx 20$ $\mu s$. The time-dependence of the
end-to-end distance shows that force-quench refolding occurs in steps.
{\bf B.} The deviation of the dihedral angles from their values in the native state as a function of  time shows  large departures from native values  of
the dihedral angles in loop region (indicated by the red strip).
Note that this strip disappears around $t\approx 300$ $\mu s$, which coincides with the formation of bonds
shown in {\bf C}. $f_B$ is the fraction of bonds in pink that  indicates that the bond is fully formed.
{\bf D.} The histograms collected from the projections of twelve stretching and force-quench refolding trajectories on the two dimensional plane characterized by the end-to-end distance ($R$) and the average correctness of dihedral angles ($\langle 1-\cos{(\phi-\phi^0)}\rangle$) around the loop region ($i=19-24$).  The scale on the right gives the density of points in the two dimensional projection.  This panel shows that the local dihedral angles are coupled to the end-to-end distance $R$, and hence extension alone is not a good reaction coordinate especially in force-quench refolding. The molecular extension $x$ is related to $R$ by $x = R -R_N$ where $R_N$ is the distance in the folded state.  

\clearpage

\newpage
\begin{figure}[ht]
\includegraphics[width=3.50in]{landscape.eps}
\caption{\label{landscape}}
\end{figure}
\clearpage
\begin{figure}[ht]
\includegraphics[width=6.50in]{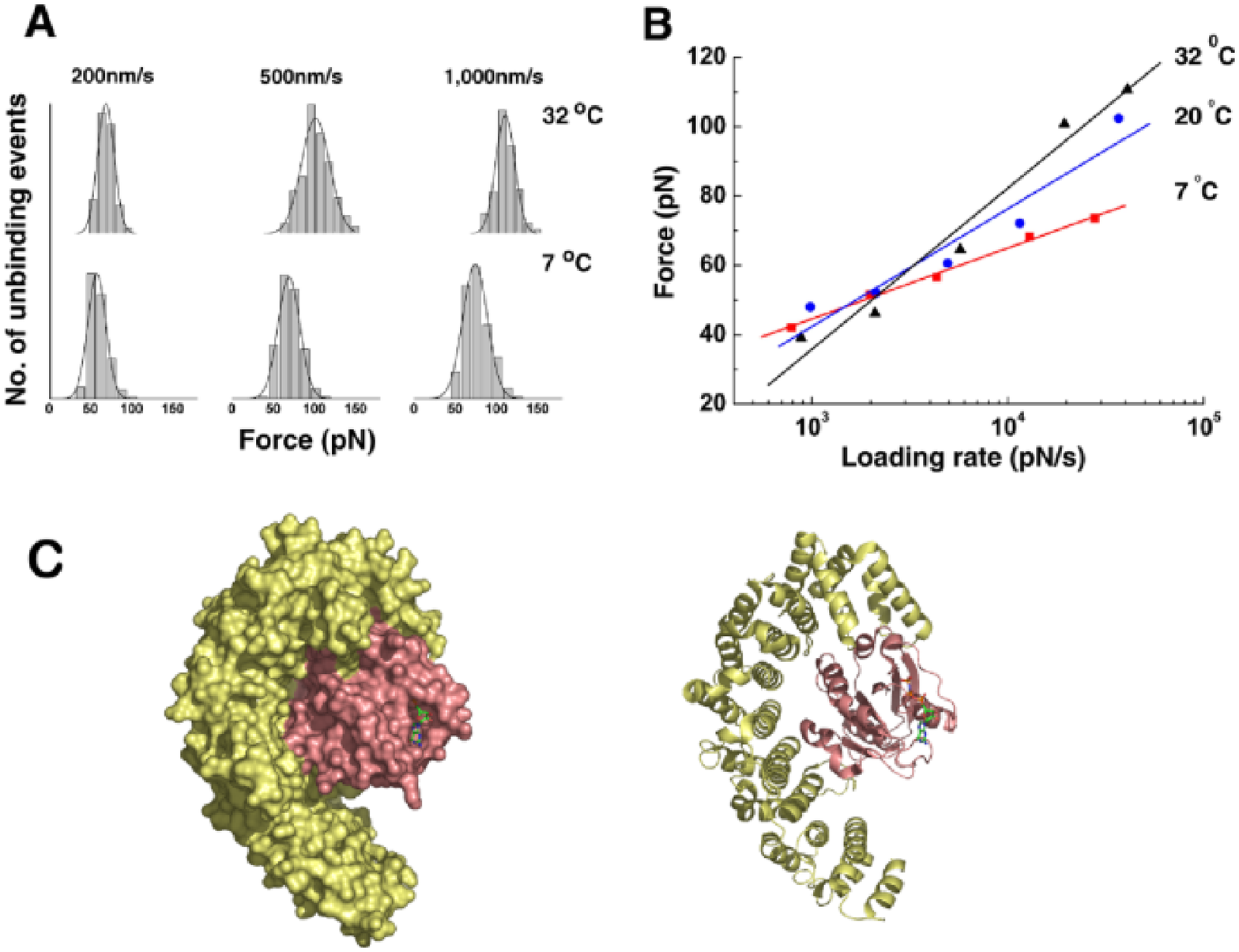}
\caption{\label{Nevo_fig}}
\end{figure}
\begin{figure}[ht]
\includegraphics[width=4.50in]{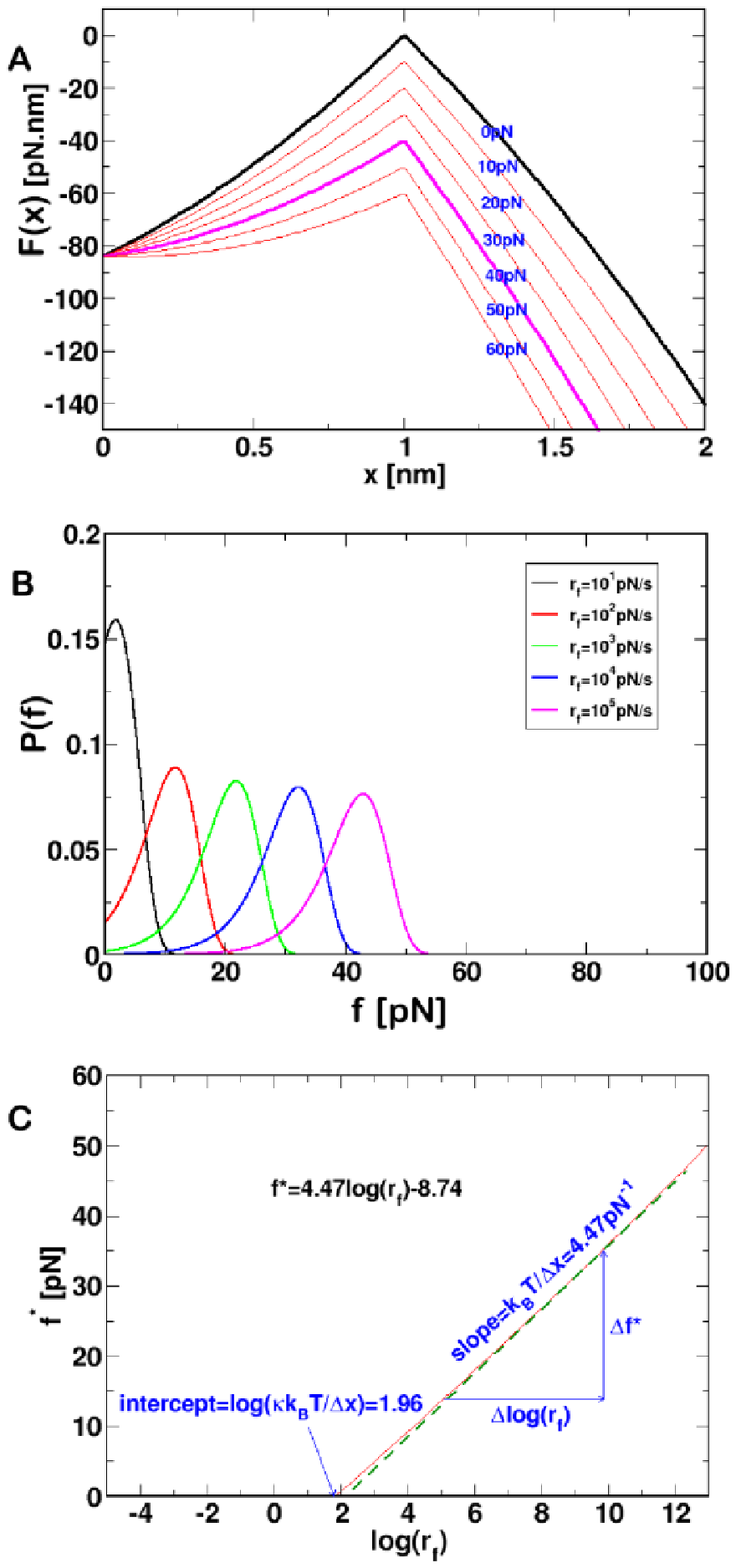}
\caption{\label{hardplot.fig}}
\end{figure}
\begin{figure}[ht]
\includegraphics[width=4.50in]{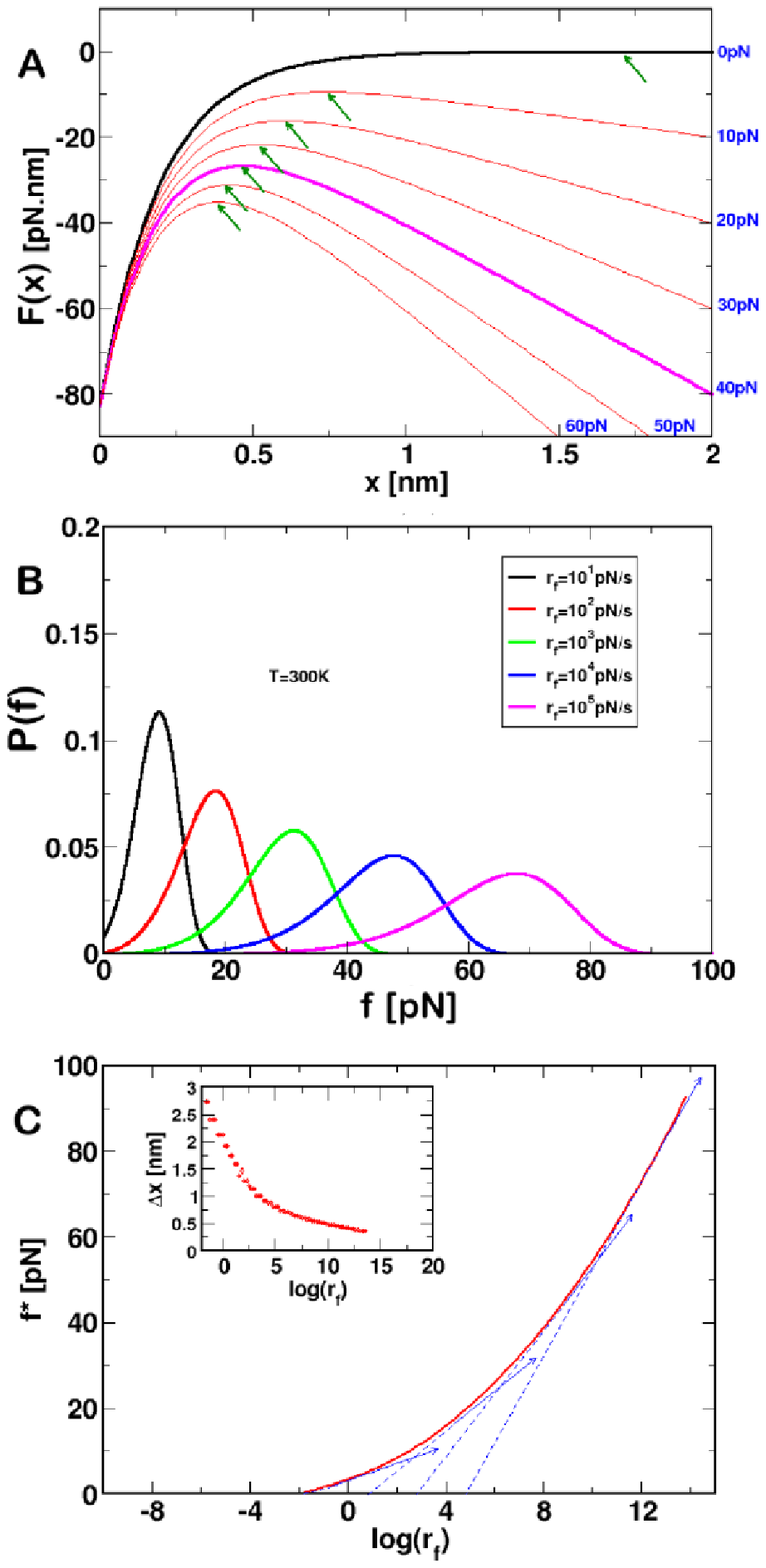}
\caption{\label{softplot.fig}}
\end{figure}

\begin{figure}[ht]
\includegraphics[width=5.50in]{hard_soft_Hammond.eps}
\caption{\label{hard_soft_Hammond.fig}}
\end{figure}
\begin{turnpage}
\begin{figure}[ht]
\includegraphics[width=9.00in]{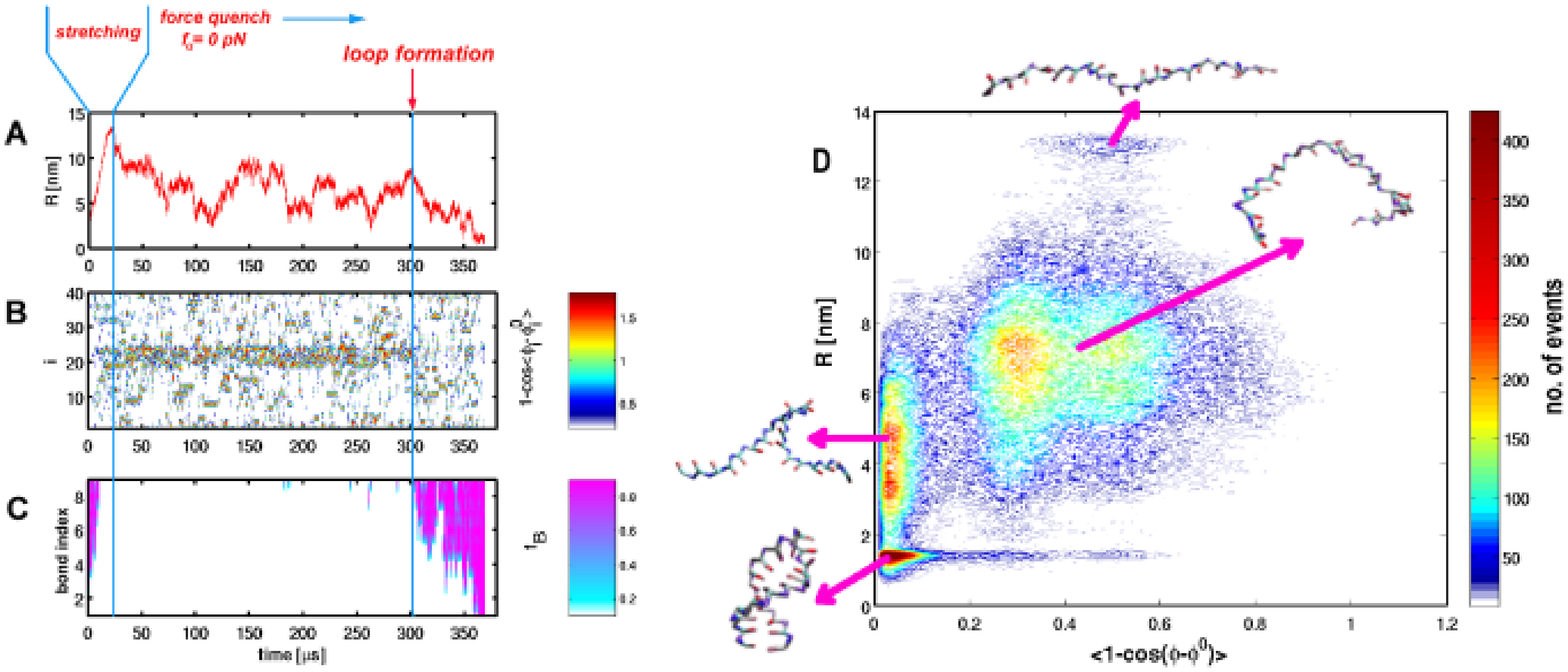}
\caption{\label{dih_R_2D_map_illust.fig}}
\end{figure}
\end{turnpage}
\end{document}